\documentclass[letterpaper,journal]{IEEEtran}
\IEEEoverridecommandlockouts

\usepackage{cite}
\usepackage{amsmath,amssymb,amsfonts}
\usepackage{algorithmic}
\usepackage{graphicx}
\usepackage{textcomp}
\usepackage{stfloats}
\usepackage{url}
\usepackage{verbatim}
\usepackage[table]{xcolor}
\usepackage{colortbl}
\usepackage{hyperref}  
\usepackage{enumitem}  
\usepackage{makecell}
\usepackage{booktabs}  

\graphicspath{{figures/}{figures/authors/}} 

\def\BibTeX{{\rm B\kern-.05em{\sc i\kern-.025em b}\kern-.08em
    T\kern-.1667em\lower.7ex\hbox{E}\kern-.125emX}}
\begin{document}

\title{Remote Awareness of Seafloor Images Collected by AUVs over Low-Bandwidth Communication Links\\
\thanks{This research was funded by the European Union's Horizon 2020 research and innovation programme under Grant Agreement No 101000858 (TechOceanS - Technology for Ocean Sensing) and by UK Research and Innovation under project reference 10110715 (OASIS - Over horizon Awareness of Seafloor Imaging Surveys).}
}

\author{
    Adrian Bodenmann \IEEEmembership{Member, IEEE},
    Cailei Liang,
    Miquel Massot-Campos,
    Samuel Simmons,
    Alexander B. Phillips,
    Alberto Consensi,
    Matthew Kingsland,
    Rashiid Sherif,
    Stan Brown,
    Adam Riese,
    Blair Thornton \IEEEmembership{Member, IEEE}
    \thanks{A. Bodenmann, C. Liang, S. Simmons, and B. Thornton are with the School of Engineering, University of Southampton, Southampton, UK.}
    \thanks{B. Thornton is also with the Institute of Industrial Science, The University of Tokyo, Tokyo, Japan.}
    \thanks{M. Massot-Campos and R. Sherif are with Ocean Infinity, Southampton, UK.}  
    \thanks{A. B. Phillips, A. Consensi and M. Kingsland are with Marine Autonomous and Robotic Systems, National Oceanography Centre, Southampton, UK.}
    \thanks{S. Brown and A. Riese are with Voyis Imaging Inc., Ontario, Canada.}
}

\maketitle

\begin{abstract}
This paper introduces a method for real-time processing and transmission of autonomous underwater vehicle (AUV) imagery over low-bandwidth communication links.
It leverages artificial intelligence (AI) techniques to identify a set of images that best represent an entire dataset, or automatically finds the most similar images to a given query image for transmission to operators. 
Combined with metadata of a larger set of images, compressed versions of the selected images can be transmitted over satellite communication links or underwater modems, and provide operators on shore with information about the type of imagery the AUV is collecting while it is still deployed.
Data from three deployments off the coast of the UK and in Gran Canaria using different AUVs and imaging systems demonstrate the method in the field.
It achieved an almost 400\,000-fold reduction in data volume compared to the raw data size, enabling transmission of data summaries of a 2-hour 47-minute-long mapping mission in just over 34 minutes over low-bandwidth satellite communication.
\end{abstract}


\bstctlcite{IEEEexample:BSTcontrol}

\section{Introduction}
Autonomous underwater vehicles (AUVs) are versatile tools, commonly used for seafloor photo surveys with applications in environmental monitoring, infrastructure inspection and search operations.
Such systems are capable of autonomously collecting large numbers of high-resolution images;
however, due to the intersection of the field of view of the camera and the illumination source (typically strobes), the distance from which such systems can operate is limited to two to tree attenuation lengths~\cite{Jaffe1988}.
With attenuation lengths for red light of typically around 4\,m in water (at 620\,nm; less for longer wavelengths)~\cite{Wozniak2007}, clear full-colour photos can be taken in water from up to 8 to 12\,m distance in ideal conditions; less in the presence of turbidity.
This limits both the area covered per photo as well as the speed at which underwater vehicles carrying these camera devices can safely navigate.
Although advances in underwater imaging technology have led to improved mapping rates in recent years, these remain significantly lower than those of mapping operations using airborne platforms or satellites.
This makes underwater mapping operations of relevant areas time-consuming and expensive, in particular due to the high costs of support vessels.
Missions in which the AUV makes longer sorties, freeing up the support vessel for other work, or shore-launched deployments without a support vessel can lower costs.
This can also reduce the environmental impact, as most of the emissions associated with AUV operations come from support vessels~\cite{Bodenmann2025}.
Developments in long-range, long-endurance AUVs~\cite{Roper2021,Phillips2023} in recent years have made this possible, leading to a shift towards more autonomous operations.
However, full image datasets can still only be downloaded after recovering the AUV as wireless long-range communication available on AUVs, including satellite and underwater acoustic communication, has data transfer rates that are much lower than data acquisition rates of imaging systems.
Longer deployment cycles without the ability to download data in between mean that AUVs collect large amounts of data before operators can assess the data quality.
Additionally, they are not aware of what data have been collected or whether items of particular interest or concern have been captured.
Without such information, decisions about whether to change mapping parameters, what areas to survey in more detail, or whether to deploy additional equipment or sensors to a location of interest can only be made with delay, or not at all (if the campaign consists of a single deployment).
Therefore, for efficiently deploying AUVs on mapping missions where several dives are carried out without recovering the vehicle, it is necessary to send some information about the collected data back to operators throughout the mission, for example, when the AUV is at the surface between dives.

Although pressure-resistant systems to communicate with the Iridium~\cite{Phillips2023,Engdahl2023,Jakuba2024,Bodenmann2025}, ARGOS~\cite{McPhail2019} or ETS-VIII~\cite{Yoshida2019} satellites and underwater acoustic modems are commonly used in subsea surveys, the low typical average transmission rates on currently available satellite communication systems that do not rely on  satellite tracking (less than 100 Bytes per second on average for Iridium short burst data (SBD) message transmission reported in \cite{Thornton2024}) are insufficient to transmit the large volumes of data from photo surveys generating Megabytes to 10s of Megabytes of data per second~\cite{Jakuba2024}.
With data acquisition rates around five orders of magnitude higher than transmission rates, AUVs would have to spend on the order of 100\,000\,s at the surface to transmit uncompressed data via satellite communication for every second spent acquiring data of the seafloor.
Standard compression with a high compression setting while preserving resolution can reduce the image file size by around two orders of magnitude (approximately 100-fold reduction in file size when applying JPEG compression with a quality setting of 30\% to an 8-bit RGB image obtained from debayering and colour balancing a 12-bit raw image acquired with the Voyis Recon LS camera system described in section~\ref{hardware}).
This still leaves a difference of around three orders of magnitude between data acquisition and data transmission rates, which more modern compression algorithms alone cannot bridge.
Also, heavily compressing and transmitting an entire image dataset consisting of thousands of images does not provide a good overview of the data being collected, but a smaller, representative selection of a few to a few tens of images combined with metadata for a larger number images can provide a more concise characterisation of the collected imagery.

Therefore, it is necessary to prioritise and compress data, and determine what type of information, such as reduced-size images, or metadata, should be sent back to operators.
Additionally, data packaging needs to be robust against the realities of satellite or underwater acoustic data transmission from AUVs, which can be slow and intermittent, and where there is no guarantee that all data packages arrive. 
Transmitting data via satellite communication requires the AUV to surface and remain there until the transmission is complete, so the duration of these should be kept short, as no seafloor imagery can be captured during this time, and loitering for extended periods of time at the surface increases risk of collisions with ships.

Transmission of data from underwater vehicles is routinely performed by Lagrangian floats, such as Argo floats~\cite{Andre2015} or gliders such as Slocum gliders~\cite{Webb2001}.
These transmit water column data, such as conductivity, temperature, and depth (CTD) profiles, along with position data, however, these are significantly smaller in terms of data volume than photo datasets.
Many AUVs also transmit basic information including their recent navigation data, including latitudes, longitudes, depths and altitudes, designed to assist operations and showing over- or underruns of any set mission parameters, and information about the AUV's status, such as battery levels~\cite{Bodenmann2025}. 
\cite{Murphy2014} demonstrated a method for compressing image, sonar and time-series data and transmitting them over acoustic relays. However, it does not analyse the image content to identify a number of representative images or send metadata of photos whose image data is not transmitted.
\cite{Girdhar2010} and \cite{Kaeli2015} introduced methods for Bayesian surprise-based clustering of images taken by an AUV.
The set of images is split into a set of clusters, where a new cluster is generated when an image is added that is sufficiently different from the existing clusters.
\cite{Kaeli2015} demonstrated how this can be used to transmit small amounts of data from an AUV to its operators, containing compressed versions of the images representing clusters and information for the other images about the clusters to which they belong.
While the method can also identify anomalies, it is sensitive to noisy data.

In this paper, we develop a method that can select and compress information about a collected set of images by an AUV to a remote operator via a low-bandwidth satellite communication link, such as Iridium  messages.
It should be able to generate data summaries of representative images for a general overview of the dataset without being influenced by moderate levels of image noise, as well as be able to identify images similar to a prompt image that the user can provide.
The size of the data should be such that it can be transmitted via satellite link using non-directional antennae within several tens of minutes at most, in order to keep the required time for an AUV to loiter at the surface at a minimum. 
Data packages should be independent of each other, so that failure to receive one data package only leads to limited loss of data, and data from other packages are still usable.


The demonstrated method, given a set of seafloor images, uses artificial intelligence (AI) based algorithms to:
\begin{enumerate}[label=\alph*)]
    \item identify a small number of images of highest interest for transmission, and
    \item compute numeric metadata from a larger set of images.
\end{enumerate}
These data are transmitted (after downsampling and compression) from an AUV to a remote operator over low-bandwidth communication links, such as satellite communication or underwater acoustic modems.
The method supports two modes of operation:

The \textbf{Summary mode} identifies clusters of similar images within a dataset and selects the most representative images from each cluster for transmission.
Metadata based on the image content including the cluster assignment, as well as position data (latitude, longitude, altitude, depth) are transmitted for a larger number of images in tabular form.
This shares similarities with the method presented in~\cite{Thornton2024}, however, the approach presented in this paper uses convolutional neural networks (CNN) or visual transformers (ViT) with a higher-dimensional latent space, which better discriminate between different types of images. The entire processing is done on the AUV, without relying on manual labelling.

In the \textbf{Query mode} the operator provides an image of interest, and the system retrieves and transmits a predefined number of the most similar images from the dataset.
It also transmits metadata of a larger number of images, containing a measure of similarity to the query image, as well as position data.

\section{Method}

\subsection{Data Processing Steps}
\begin{figure}[]
    \centering
    \includegraphics[width=.98\columnwidth]{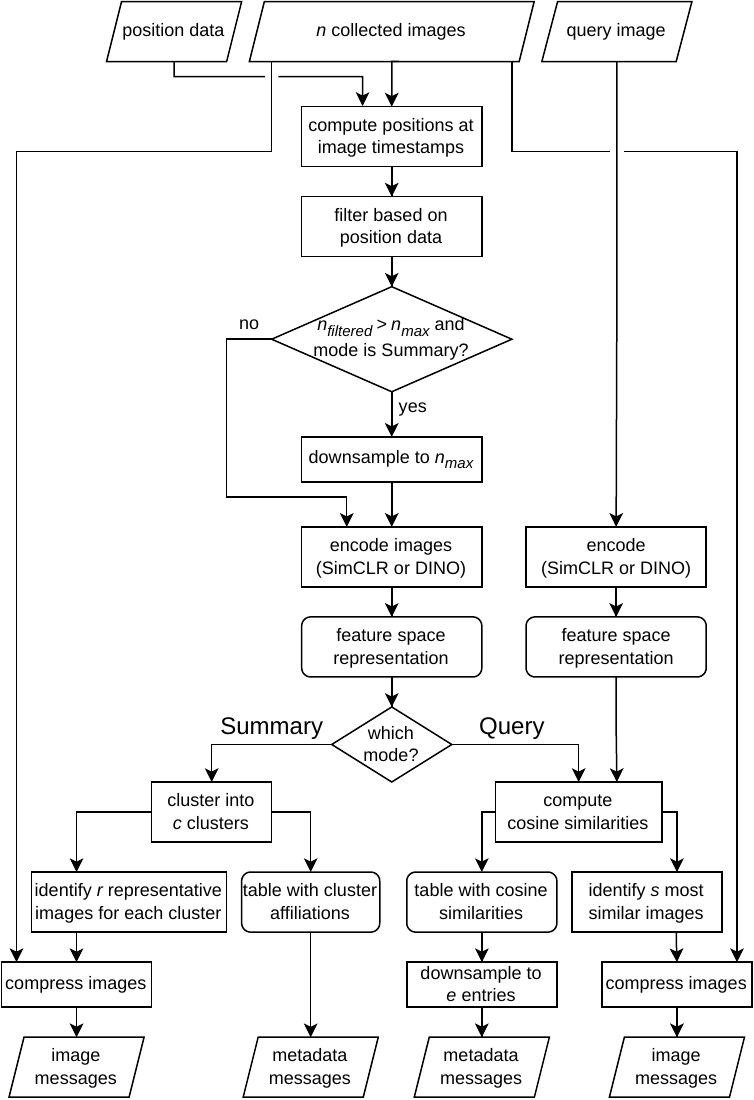}
    \caption{
        Flow diagram showing the processing steps of the data selection and compression algorithm.
    }
    \label{fig:flow_diagram}
\end{figure}

The proposed algorithm requires images from a seafloor mapping device and position data, and, when using the query mode, a prompt image.
These data are processed with an algorithm, which includes steps for data compression, transmission, and visualisation.
The algorithm whose major steps are shown in the flow diagram in Fig.~\ref{fig:flow_diagram} is invoked at the end of the mapping mission, at a specific time when sufficient images are available, or when requested by the remote operator.
When running the summary mode, the user can select the maximum number of images $n_{max}$ on which the processing is applied, the number of clusters $c$ and the number of representative images per cluster $r$ that should be identified.
When running the query mode, the user provides the query image and selects the number $s$ of most similar images to be transmitted and the number $e$ of images whose metadata is to be sent.

\subsection{Data Preprocessing and Encoding into Feature Space}
In the first step of data processing, location data for each image are inferred from the vehicle position data by interpolating to the image timestamps.
If the user requests a summary and there are more images in the dataset than a predefined threshold $n_{max}$, a random subsample of $n_{max}$ images is selected for further processing, otherwise all images are used.
If the mapping system does not already provide colour-corrected images, colour correction is applied using the method described in~\cite{Massot-Campos2023}, and the images are downsampled.
The images are then encoded, either using SimCLR (simple framework
for contrastive learning of visual representations) \cite{Chen2020} or DINO (self-distillation with no labels) \cite{Caron2021}, into a 128-dimensional latent space, in case of SimCLR, or a 768-dimensional latent space, in the case of DINO.
t-SNE (t-distributed stochastic neighbor embedding) \cite{VanDerMaaten2008} dimensional reduction is subsequently applied to project the latent space into a 2D space, which is later used for visualisation.

\subsection{Identifying Images for Transmission}
If the user requests a summary, hierarchical k-means clustering is then applied on the full latent space representation to group similar images together.
The top-level clusters, whose number $c$ can be defined by the user, represent the main categories of the images.
To identify a number $r$ of images representative of each category, the second layer of clustering is applied within each top-level cluster, and the images closest to the centroids of these subclusters are selected for transmission.
This way, a total of $c \cdot r$ images are selected.
On the other hand, if the users request a query, the same initial steps are followed, but instead of clustering, the latent space representation is used to compute a measure of similarity for each image.
For this, the cosine similarity between the query image and all other images in the latent space is computed, and the top $s$ most similar images are selected.

For both types of request, the selected images are then downsampled and compressed using Better Portable Graphics (BPG)~\cite{Bellard2018} to within a target size.
This is done iteratively, where the file size of the compressed images is checked against the target size, and the compression parameter is increased until the file size is smaller than the target size, or the maximum compression parameter is reached, in which case the file is split into multiple chunks of the prescribed target size.

\subsection{Metadata Packaging into Compact Binary Files}
Metadata of the sampled images is compressed using Dynamic Compact Control Language (DCCL)~\cite{Schneider2015}, which packages data into compact binary datagrams, designed for efficient transmission over low-bandwidth networks.
The metadata includes the image identifier, latitude, longitude, depth, altitude, and t-SNE projection values.
For summaries, the cluster number is also included, and for queries, the similarity score of the images is included.
The metadata of multiple images are combined into a file, where the number is adjusted so as to yield binary files not exceeding the maximum size the transport layer (e.g. satellite communication message or acoustic modem) can handle.
For summaries, metadata for all processed images are sent.
For queries, metadata of the top $e/2$ most similar images, plus a random sample of the remaining $e/2$ images are sent.
The resulting binary files from the DCCL-compressed metadata, as well as the BPG-compressed bitmap data are saved to a defined directory.

\subsection{Data Transmission}
The binary files are then either accessed directly by the software handling the data transmission, or, if the system handling the communication between the AUV and the shore or the ship does not have direct file access to the mapping system, the files can also be sent over RS-232 serial connection as ASCII in hexadecimal representation.
To transmit the files to shore or the support vessel, the AUV needs to surface and loiter at the surface until the transmission is complete if using satellite communication or remain within range of the acoustic modem if using underwater acoustics communication.
The time required for the transmission depends on the amount of data that need to be sent, which  depends on the parameters chosen for the summary or query, the communication method and the quality of the connection.
The satellite connection quality depends on the number and position of communication satellites that are in view of the AUV, as well as the washover from waves covering the antenna.
For acoustic underwater communication, it depends on environmental conditions, including changes in water temperature or salinity, and background noise.
As a result, the transmission will take longer than the duration calculated based on the data size and the theoretical communication bandwidth, which should be factored in when planning the mission timeline.

\subsection{Data Visualisation}
\begin{figure*}[]
    \centering
    \includegraphics[width=0.79\textwidth]{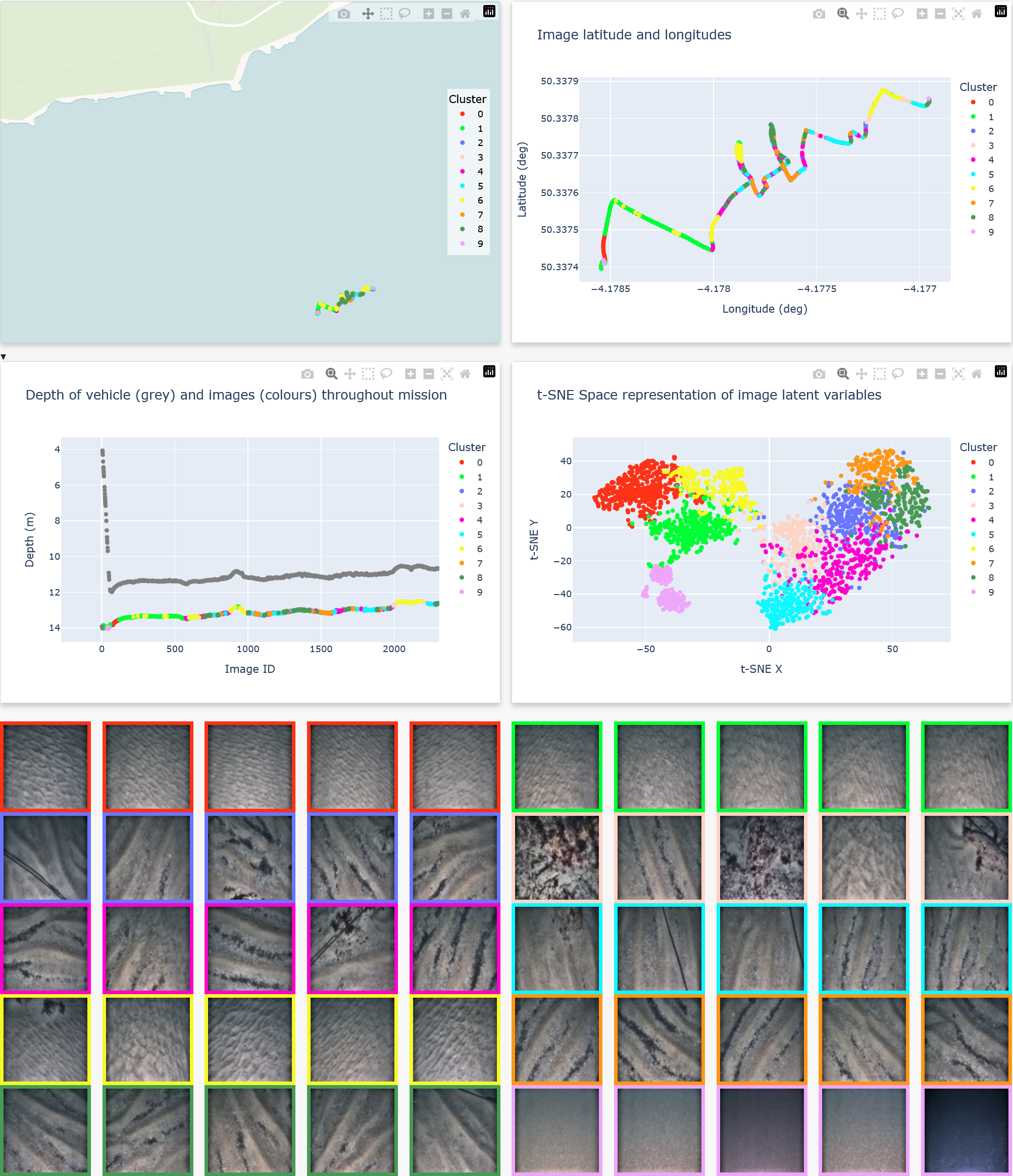}
    \caption{
        Visualisation of summary data of a dataset collected in Cawsand Bay, off Plymouth, UK.
        The map and time series plot show the location of the images and their acquisition depth and altitude.
        The t-SNE space representation shows the 10 latent space clusters identified with k-means.
        The colour of the data points and frames around the thumbnails indicate the cluster the images and data points belong to.
    }
    \label{fig:remote_summary}
\end{figure*}

\begin{figure*}[]
    \centering
    \includegraphics[width=0.785\textwidth]{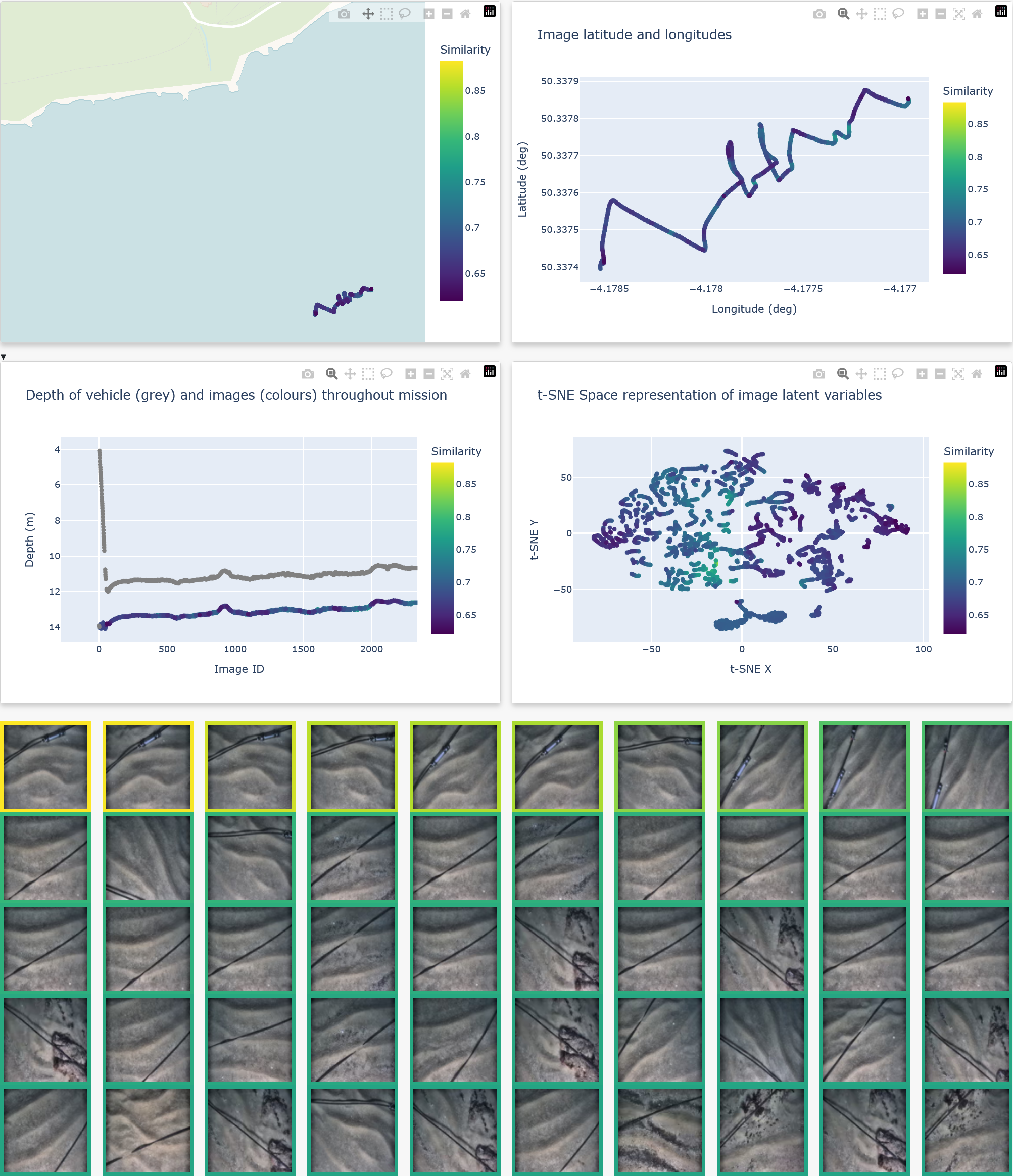}
    \caption{
        Visualisation of query results applied on the same dataset as used in Fig.~\ref{fig:remote_summary}.
        The image of the cable with a mock-up repeater shown in Fig.~\ref{fig:query_analysis_result} was used as query image and the algorithm was run with the DINO encoder.
        The thumbnails of the identified images for transmission at the bottom are in order of similarity.
        As there are only 12 images showing the mock-up repeater (full or partially) in the entire dataset, the algorithm also returned images that contain cable and similar types of substrate as the query image, which are the next-best matches in the dataset.
    }
    \label{fig:remote_query}
\end{figure*}

After the files are received, they are visualised in a web browser via a page where the compressed image and metadata files can be supplied and loaded.
The backend runs a server that, when a transmitted file is supplied via the browser page, unpacks the DCCL-compressed metadata and decodes the BPG-compressed images and displays them.
Plots show the colour-coded clustering or similarity scores on a map, in the t-SNE projection and as time sequence plots showing the depth and acquisition altitude, as shown in Figs.~\ref{fig:remote_summary} and~\ref{fig:remote_query}.

\section{Results}

\subsection{Hardware Setup}
\label{hardware}

\begin{figure}
    \centering
    \includegraphics[width=.87\columnwidth]{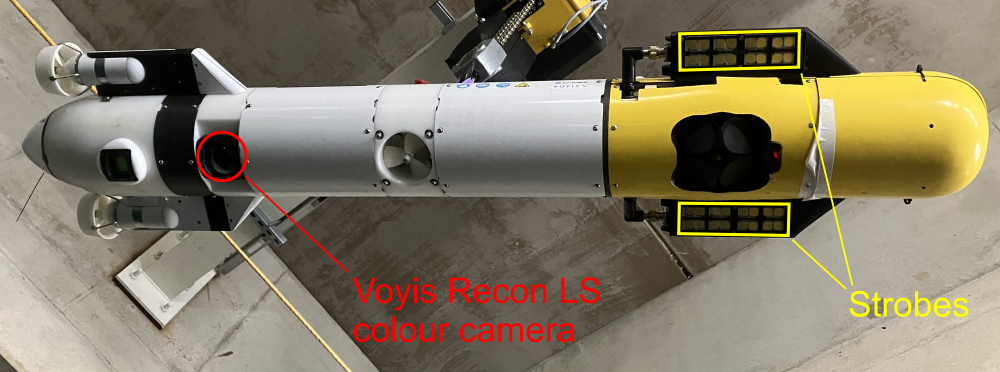}
    \caption{
        \label{fig:smarty200_voyis}
        Smarty200 AUV equipped with Voyis Recon LS camera system.
    }
    \vspace{5mm}
    \includegraphics[width=.87\columnwidth]{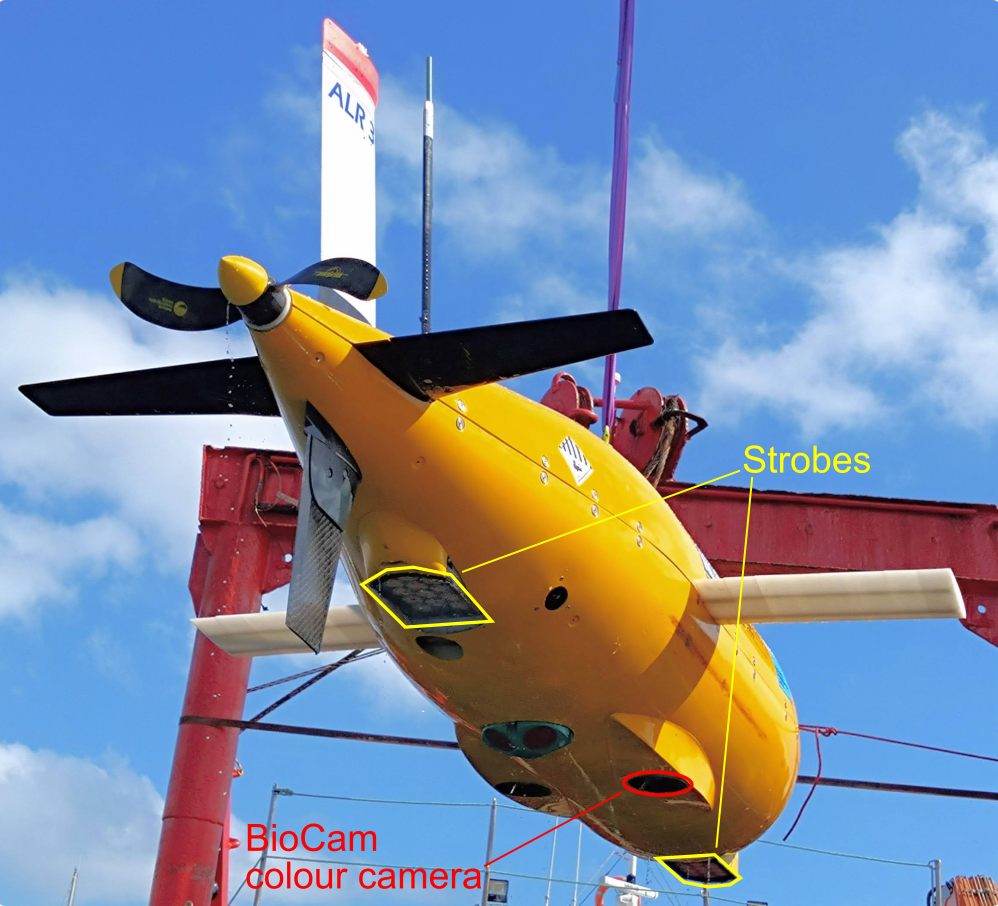}
    \caption{
        \label{fig:alr_biocam}
        6000\,m rated version of ALR equipped with the \mbox{BioCam} camera system.
    }
    \vspace{5mm}
    \includegraphics[width=.87\columnwidth]{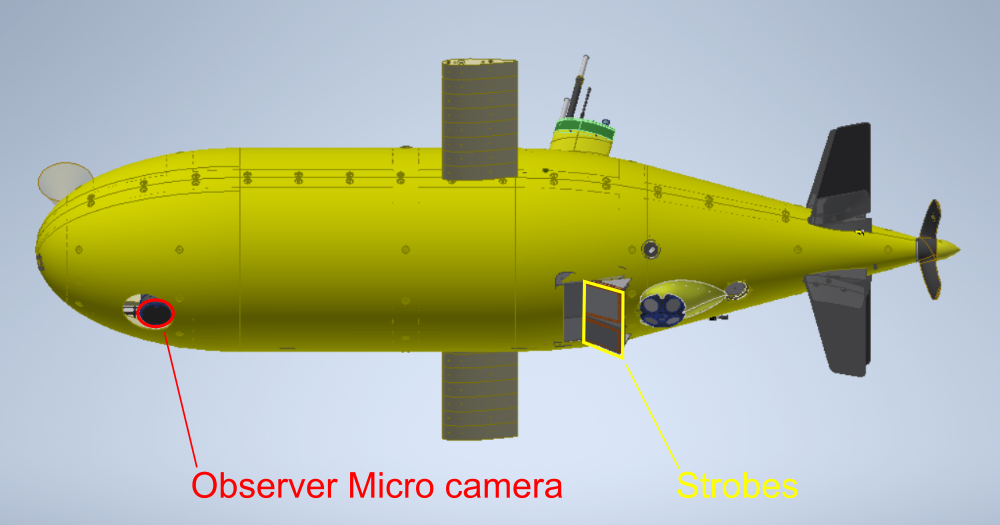}
    \caption{
        \label{fig:alr_voyis}
        1500\,m rated version of ALR equipped with the Voyis Observer Micro camera system.
    }
\end{figure}

The method has been deployed on different combinations of AUVs and seafloor mapping devices.
The data presented in this paper has been collected with three setups:
the ``Smarty200'' Iqua Robotics Sparus II AUV equipped with a Voyis Recon LS camera system~\cite{Thornton2024} shown in Fig.~\ref{fig:smarty200_voyis}, 
ALR3, which is a 6000\,m rated version of the Autosub Long Range AUV (ALR; a.k.a. ``Boaty McBoatface'')~\cite{Roper2021}, where the \mbox{BioCam} camera system had been integrated as explained in~\cite{Bodenmann2025} and shown in Fig.~\ref{fig:alr_biocam}, and 
ALR8, which is a 1500\,m rated version of ALR~\cite{Phillips2023}, equipped with the Voyis Observer Micro camera system shown in Fig.~\ref{fig:alr_voyis}.
Details of the three setups are provided in table~\ref{tab:hardware}.

\begin{table*}[h!]
	\begin{center}
		\caption{
			AUVs and mapping systems used for demonstrating the remote awareness of seafloor images algorithm.
		}
		\vspace{-0.2cm}
		\label{tab:hardware}
		\rowcolors{0}{gray!10}{white}
		\begin{tabular}{l|ccc}
			\toprule
			AUV                                      & Smarty200 (Iqua Robotics Sparus II)             &  Autosub Long Range 3 (ALR3)                        & Autosub Long Range 8 (ALR8)                      \\
			Mapping system                           & Voyis Recon LS                                  &  BioCam                                             & Voyis Observer Micro                             \\
			\midrule  
			AUV range and autonomy              & 12\,km / 12\,h                                  &  1800\,km / 3 months                                & 6000\,km / 6 months                              \\
			AUV depth rating                         & 200\,m                                          &  6000\,m                                            & 1500\,m                                          \\
			Image resolution, size, frame rate      & 12\,MP / 24\,MB @ 1\,Hz                         &  5\,MP / 11\,MB @ 0.33\,Hz                          & 12\,MP / 24\,MB @ 1\,Hz                          \\
			Illumination                             & 2 LED strobes                                   &  2 LED strobes                                      & 2 LED strobes                                    \\
			\Gape[0pt][2pt]{\makecell[l]{Communication \\ mapping system - AUV}}       & Ethernet                                       &  \makecell{RS-232 (during missions), \\ Ethernet (setup, data extraction)} & Ethernet                                         \\
			\makecell[l]{Communication with operators \\ short-range / over-horizon} & \makecell{Wi-Fi / Iridium SBD \\ 340 bytes per message} &  \makecell{Wi-Fi / Iridium SBD \\ 1960 bytes per message}   & \makecell{Wi-Fi / Iridium SBD \\ 1960 bytes per message} \\
			\bottomrule
		\end{tabular}
	\end{center}
\end{table*}

\subsection{Data Collection}

\subsubsection{Infrastructure Survey Demonstration}
During the four-day-long EE25-01 campaign out of Plymouth, UK, in March 2025, a survey of a subsea communication cable was demonstrated with Smarty200 and the Voyis Recon LS camera system~\cite{Djauhari2026}.
A 100\,m long section of fibre optic subsea communication cable along with a mock-up signal repeater in the shape of a metal cylinder was temporarily laid in Cawsand Bay off Plymouth at approximately 15\,m depth and mapped using the Smarty200 AUV over four consecutive days.
Operations for laying and recovering the cable, and for deploying the AUV were conducted from the coastal survey vessel \textit{Echo Explorer}.
During multiple dives, the AUV gathered images of the cable, the mock-up signal repeater and the surrounding seafloor using waypoint tracking as well as using an algorithm for autonomously following the cable.
The survey vessel remained within \mbox{Wi-Fi} range of the AUV, for monitoring and control purposes.
Time allowing, the remote summaries or queries were requested upon surfacing of the AUV over the \mbox{Wi-Fi} connection.
Due to an electrical issue, the AUV computer was unable to interface with the Iridium satellite modem, however, the output from the remote summary and query could be transmitted via \mbox{Wi-Fi}.
Results from two remote summaries and six queries were successfully transmitted and visualised during the campaign.
Summary and query results are shown in figures \ref{fig:remote_summary} and \ref{fig:remote_query}.
As the thumbnails in Fig.~\ref{fig:remote_summary} show, clusters 0, 1 and 6 are of a very similar type, which a human observer would likely classify as the same class.
This is also reflected by the t-SNE representation, where the metadata points of the three clusters are grouped together.
Classes that represent many images are often split into multiple clusters, as seen where, which is the reason why the number of clusters specified (10 in this example) needs to be larger than the number of classes in the image dataset.
This also allows to better accommodate for images of the same type of seafloor acquired from a different altitude or different lighting, as is the case for cluster number 9, which is the same type of seafloor as clusters 0, 1 and 6, but acquired from higher altitude.
Further analysis of the results and their repeatability is described in section \ref{sec:analysis}.

\subsubsection{Environmental Monitoring Demonstration off Gran Canaria}
In March 2024, an earlier version of the summary method was demonstrated during the techoceans2024 campaign in Gran Canaria, Spain, using ALR3 and \mbox{BioCam} ~\cite{Plocan2024}.
The deployment was shore-based, where the only boat used was a rigid hull inflatable boat (RHIB) to tow the AUV in and out of the harbour.
\mbox{BioCam} collected two datasets in which strobed images were acquired while mapping the seafloor at a depth between 30\,m and 40\,m in repeat grid surveys from 10\,m and 5\,m altitude.

During the first survey, 3326 colour images were collected by \mbox{BioCam} during 2 hours and 47 minutes with a total file size of 34.3\,GB.
1500 of these were sampled for the remote summary, and after altitude filtering, 1477 images remained for further processing.
As \mbox{BioCam} stores images in raw format, the images were then colour corrected and downsampled.
The summary algorithm was set up to split the data into 16 classes, with one image each to be transmitted via satellite communication.
The 16 images and metadata of the 1477 images were compressed into 54 Iridium binary files, totalling 90.5\,KB of data.
These were transmitted from \mbox{BioCam} to ALR3 over RS-232, and from ALR3 to shore as Iridium SBD messages, with the total transmission taking 34 minutes and 24 seconds, corresponding to a rate of 1.57 messages per minute or 44 bytes per second on average.

\subsubsection{Over-Horizon Piloted Environmental Monitoring Demonstration off Shetland}  
During the OASIS ALR8 Shetland Trials in June 2025 ALR8 and the Voyis Observer Micro camera system were deployed out of Lerwick in Shetland, UK.  
ALR8 was deployed on a 3-day shore-launched mission covering 151\,km, piloted over the horizon from shore using Iridium satellite communication.
During the deployment, the Voyis Observer Micro camera system was active for approximately 10.5 hours, collecting 37285 colour images at a depth of between 76\,m and 82\,m depth from 4\,m altitude across multiple dives.

A modified version of the image summary method classed the collected images into 493 classifications of 3 distinct classes and compressed their metadata into 9 binary files with an average file size of 1.7\,kB and a total of 13\,kB. These were transmitted by the ALR8 over its Iridium satellite communication link with an average latency of 8 minutes (standard deviation of 11 minutes), from transmission to being available for visualisation on the Command and Control (C2) interface.

\subsection{Analysis of Results vs. Ground Truth and Repeatability}
\label{sec:analysis}
The performance of the summary and query algorithms were assessed using the data from the EE25-01 campaign.
Data were reprocessed offline to gauge aspects of alignment with ground truth, repeatability and dependence on the encoder model.
The image-type ground truth was generated by manually labelling all images in a dataset of 4292 images.
The summary and query algorithms were run using both the SimCLR and DINO models for encoding.
The SimCLR model was trained using an image dataset collected at the same location the previous year, as SimCLR models trained with data similar to the target tend to perform better~\cite{Liang2025LRSSL}.
For DINO, the generic model pretrained on the ImageNet-1k dataset was used~\cite{Caron2021}, due to the high computational resources required to train DINO models.

\subsubsection{Summary Performance}
In order to assess how well the generated summaries represent the images of the entire dataset, the summary algorithm was run multiple times using both models to encode the images, and the alignment of the clusters with the ground truth classes was evaluated.
Clusters identified by the summary algorithm should ideally contain images from a single ground truth class to be representative of that class, though multiple clusters can represent the same ground truth class.
If this is the case, clusters are representative of a group of images that a human observer would also judge to be similar, and so can be meaningfully represented by a transmitted image.
The summary algorithm was run using a subsample of 3000 images, with 10 top-level clusters and 5 representative images per cluster, using both the SimCLR and DINO models for encoding.
100 repeat runs were performed each to assess the stability and repeatability of the results.
For each run the alignment of the clusters with the ground truth classes was evaluated, by comparing for each image into what cluster it was assigned, and what ground truth class it belonged to.
This was done using all 3000 sampled images, as well as only considering the 50 representative images that were selected for transmission.
The cluster vs. ground truth class matrices in tables \ref{tab:cluster_class_matrices_simclr_3000} to \ref{tab:cluster_class_matrices_dino_50} show the results of this evaluation for 5 of the 100 repeat runs each.
Six image classes were considered for this (``Water column'', ``Mud'', ``Sand'', ``Rock'', ``Sand and rock'', and ``Sand and mud'').

\begin{figure}
    \centering
    \includegraphics[width=0.85\columnwidth]{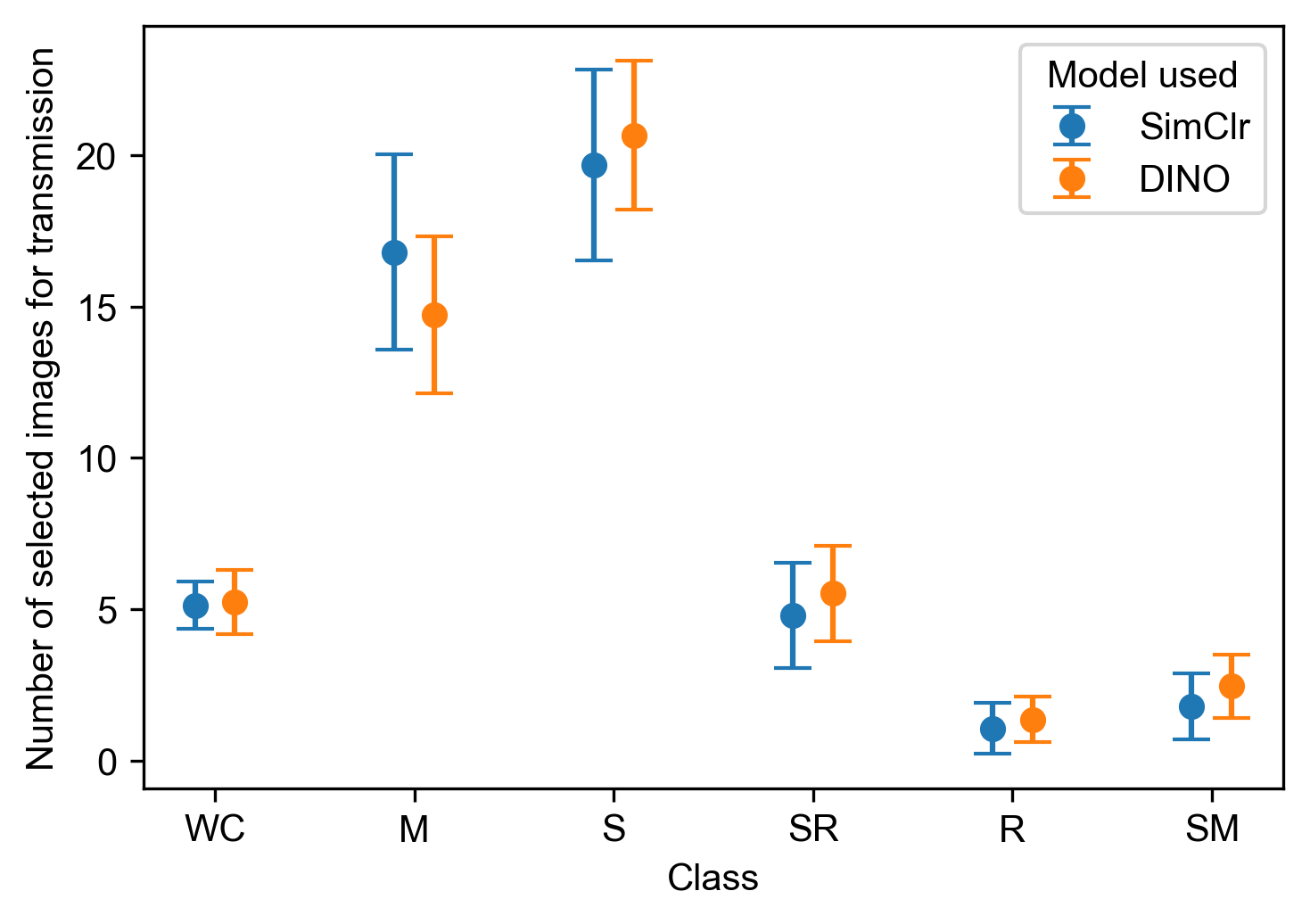}
    \caption{
        Average number and standard deviation of the number of images selected for transmission per class for 100 repeat runs of remote summary.
    }
    \label{fig:statistics_of_transmitted_classes_in_summaries}
\end{figure}

Generating the clusters of images as part of the summary algorithm involves various sources of randomness. 
This includes the random subsampling of images from the full dataset, as well as the initialisation of k-means clustering, which leads to different cluster compositions for different runs.
This also leads to variations in the cluster-class matrices, as shown in tables \ref{tab:cluster_class_matrices_simclr_3000} to \ref{tab:cluster_class_matrices_dino_50}.
The results show that the clustering behavior is consistent across runs. 
While the number of clusters representing a ground truth class can vary (e.g. ``Mud'' class represented by 3 clusters in the lef-most matrices of tables \ref{tab:cluster_class_matrices_simclr_3000} and \ref{tab:cluster_class_matrices_simclr_50}, and 4 clusters in the matrices second from the left), and some clusters straddle multiple ground truth classes, they generally represent ground truth classes well.
Part of the reason why clusters contain images from multiple ground truth classes is due to the inavoidable overlap already in the ground truth (e.g. ``Sand and rock'' class containing images of both sand and rock, which are also classes on their own).
For the same reason classes of mixed image types were not picked up in the clustering process.
Also, classes with very few images (e.g. ``Rock'') tend not to be well represented in the clusters as there are not enough images to form a distinct cluster.
The clusters of summaries based on the encoding with the DINO model show a slightly better alignment with the ground truth classes than those based on the SimCLR model, with the maxium number of images falling into a single class being higher for DINO in 4 out of the 5 runs shown in tables \ref{tab:cluster_class_matrices_simclr_3000} and \ref{tab:cluster_class_matrices_dino_3000}.
While the results differ between runs due to the randomness in sampling and the clustering process, the overall trends remain consistent for both models.
This is reflected by a relatively low variability in the number of images selected for transmission per class, as shown in Fig.~\ref{fig:statistics_of_transmitted_classes_in_summaries}.

\addtolength{\tabcolsep}{-0.52em}
\begin{table*}
    \caption{
        Cluster vs. ground truth class matrices for 5 repeat runs of remote summaries of the 3000 random sampled images, encoded with the SimCLR model.
        The ground truth classes considered for this were ``Water column'' (WC), ``Mud'' (M), ``Sand'' (S), ``Sand and rock'' (SR), ``Rock'' (R) and ``Sand and mud'' (SM).
        The cluster number (Cl) is the identifier for each cluster and carries no intrinsic meaning.
    }
    \parbox{35mm}{
        \centering
\begin{tabular}{rrrrrrr}
Cl & WC & M & S & SR & R & SM \\
9 & \cellcolor[HTML]{17cbe4} \textcolor[HTML]{000000}{197} & \cellcolor[HTML]{7a09ff} \textcolor[HTML]{f1f1f1}{10} & \cellcolor[HTML]{000000} \textcolor[HTML]{f1f1f1}{0} & \cellcolor[HTML]{000000} \textcolor[HTML]{f1f1f1}{0} & \cellcolor[HTML]{000000} \textcolor[HTML]{f1f1f1}{0} & \cellcolor[HTML]{000000} \textcolor[HTML]{f1f1f1}{0} \\
1 & \cellcolor[HTML]{7019ff} \textcolor[HTML]{f1f1f1}{23} & \cellcolor[HTML]{8bfeae} \textcolor[HTML]{000000}{349} & \cellcolor[HTML]{000000} \textcolor[HTML]{f1f1f1}{0} & \cellcolor[HTML]{000000} \textcolor[HTML]{f1f1f1}{0} & \cellcolor[HTML]{8000ff} \textcolor[HTML]{f1f1f1}{2} & \cellcolor[HTML]{780dff} \textcolor[HTML]{f1f1f1}{12} \\
0 & \cellcolor[HTML]{8000ff} \textcolor[HTML]{f1f1f1}{2} & \cellcolor[HTML]{c8e688} \textcolor[HTML]{000000}{430} & \cellcolor[HTML]{000000} \textcolor[HTML]{f1f1f1}{0} & \cellcolor[HTML]{000000} \textcolor[HTML]{f1f1f1}{0} & \cellcolor[HTML]{000000} \textcolor[HTML]{f1f1f1}{0} & \cellcolor[HTML]{8000ff} \textcolor[HTML]{f1f1f1}{1} \\
5 & \cellcolor[HTML]{000000} \textcolor[HTML]{f1f1f1}{0} & \cellcolor[HTML]{52f5cb} \textcolor[HTML]{000000}{278} & \cellcolor[HTML]{000000} \textcolor[HTML]{f1f1f1}{0} & \cellcolor[HTML]{000000} \textcolor[HTML]{f1f1f1}{0} & \cellcolor[HTML]{8000ff} \textcolor[HTML]{f1f1f1}{3} & \cellcolor[HTML]{780dff} \textcolor[HTML]{f1f1f1}{13} \\
3 & \cellcolor[HTML]{000000} \textcolor[HTML]{f1f1f1}{0} & \cellcolor[HTML]{4856fb} \textcolor[HTML]{f1f1f1}{75} & \cellcolor[HTML]{5c38fd} \textcolor[HTML]{f1f1f1}{50} & \cellcolor[HTML]{218cf4} \textcolor[HTML]{f1f1f1}{126} & \cellcolor[HTML]{5c38fd} \textcolor[HTML]{f1f1f1}{50} & \cellcolor[HTML]{780dff} \textcolor[HTML]{f1f1f1}{12} \\
\end{tabular}

    }
    \hfill
    \parbox{35mm}{
        \centering
        \begin{tabular}{rrrrrrr}
Cl & WC & M & S & SR & R & SM \\
9 & \cellcolor[HTML]{1acfe3} \textcolor[HTML]{000000}{204} & \cellcolor[HTML]{7e03ff} \textcolor[HTML]{f1f1f1}{4} & \cellcolor[HTML]{000000} \textcolor[HTML]{f1f1f1}{0} & \cellcolor[HTML]{000000} \textcolor[HTML]{f1f1f1}{0} & \cellcolor[HTML]{000000} \textcolor[HTML]{f1f1f1}{0} & \cellcolor[HTML]{000000} \textcolor[HTML]{f1f1f1}{0} \\
8 & \cellcolor[HTML]{6e1cff} \textcolor[HTML]{f1f1f1}{26} & \cellcolor[HTML]{12c8e6} \textcolor[HTML]{000000}{193} & \cellcolor[HTML]{000000} \textcolor[HTML]{f1f1f1}{0} & \cellcolor[HTML]{000000} \textcolor[HTML]{f1f1f1}{0} & \cellcolor[HTML]{000000} \textcolor[HTML]{f1f1f1}{0} & \cellcolor[HTML]{8000ff} \textcolor[HTML]{f1f1f1}{3} \\
3 & \cellcolor[HTML]{7c06ff} \textcolor[HTML]{f1f1f1}{8} & \cellcolor[HTML]{48f1d0} \textcolor[HTML]{000000}{264} & \cellcolor[HTML]{8000ff} \textcolor[HTML]{f1f1f1}{1} & \cellcolor[HTML]{000000} \textcolor[HTML]{f1f1f1}{0} & \cellcolor[HTML]{7e03ff} \textcolor[HTML]{f1f1f1}{5} & \cellcolor[HTML]{780dff} \textcolor[HTML]{f1f1f1}{14} \\
2 & \cellcolor[HTML]{000000} \textcolor[HTML]{f1f1f1}{0} & \cellcolor[HTML]{9efaa2} \textcolor[HTML]{000000}{375} & \cellcolor[HTML]{000000} \textcolor[HTML]{f1f1f1}{0} & \cellcolor[HTML]{000000} \textcolor[HTML]{f1f1f1}{0} & \cellcolor[HTML]{000000} \textcolor[HTML]{f1f1f1}{0} & \cellcolor[HTML]{000000} \textcolor[HTML]{f1f1f1}{0} \\
5 & \cellcolor[HTML]{000000} \textcolor[HTML]{f1f1f1}{0} & \cellcolor[HTML]{30e1da} \textcolor[HTML]{000000}{231} & \cellcolor[HTML]{000000} \textcolor[HTML]{f1f1f1}{0} & \cellcolor[HTML]{000000} \textcolor[HTML]{f1f1f1}{0} & \cellcolor[HTML]{8000ff} \textcolor[HTML]{f1f1f1}{1} & \cellcolor[HTML]{7a09ff} \textcolor[HTML]{f1f1f1}{9} \\
\end{tabular}

    }
    \hfill
    \parbox{35mm}{
        \centering
        \begin{tabular}{rrrrrrr}
Cl & WC & M & S & SR & R & SM \\
6 & \cellcolor[HTML]{1fd3e1} \textcolor[HTML]{000000}{209} & \cellcolor[HTML]{7c06ff} \textcolor[HTML]{f1f1f1}{7} & \cellcolor[HTML]{000000} \textcolor[HTML]{f1f1f1}{0} & \cellcolor[HTML]{000000} \textcolor[HTML]{f1f1f1}{0} & \cellcolor[HTML]{000000} \textcolor[HTML]{f1f1f1}{0} & \cellcolor[HTML]{000000} \textcolor[HTML]{f1f1f1}{0} \\
0 & \cellcolor[HTML]{7413ff} \textcolor[HTML]{f1f1f1}{18} & \cellcolor[HTML]{ff6b37} \textcolor[HTML]{f1f1f1}{576} & \cellcolor[HTML]{000000} \textcolor[HTML]{f1f1f1}{0} & \cellcolor[HTML]{000000} \textcolor[HTML]{f1f1f1}{0} & \cellcolor[HTML]{7e03ff} \textcolor[HTML]{f1f1f1}{5} & \cellcolor[HTML]{7019ff} \textcolor[HTML]{f1f1f1}{23} \\
1 & \cellcolor[HTML]{8000ff} \textcolor[HTML]{f1f1f1}{1} & \cellcolor[HTML]{dbd87b} \textcolor[HTML]{000000}{455} & \cellcolor[HTML]{000000} \textcolor[HTML]{f1f1f1}{0} & \cellcolor[HTML]{000000} \textcolor[HTML]{f1f1f1}{0} & \cellcolor[HTML]{000000} \textcolor[HTML]{f1f1f1}{0} & \cellcolor[HTML]{000000} \textcolor[HTML]{f1f1f1}{0} \\
8 & \cellcolor[HTML]{000000} \textcolor[HTML]{f1f1f1}{0} & \cellcolor[HTML]{3e65fa} \textcolor[HTML]{f1f1f1}{88} & \cellcolor[HTML]{504afc} \textcolor[HTML]{f1f1f1}{66} & \cellcolor[HTML]{7216ff} \textcolor[HTML]{f1f1f1}{20} & \cellcolor[HTML]{7a09ff} \textcolor[HTML]{f1f1f1}{9} & \cellcolor[HTML]{6c1fff} \textcolor[HTML]{f1f1f1}{29} \\
3 & \cellcolor[HTML]{000000} \textcolor[HTML]{f1f1f1}{0} & \cellcolor[HTML]{7216ff} \textcolor[HTML]{f1f1f1}{21} & \cellcolor[HTML]{18cde4} \textcolor[HTML]{000000}{202} & \cellcolor[HTML]{642cfe} \textcolor[HTML]{f1f1f1}{39} & \cellcolor[HTML]{000000} \textcolor[HTML]{f1f1f1}{0} & \cellcolor[HTML]{7216ff} \textcolor[HTML]{f1f1f1}{21} \\
\end{tabular}

    }
    \hfill
    \parbox{35mm}{
        \centering
        \begin{tabular}{rrrrrrr}
Cl & WC & M & S & SR & R & SM \\
9 & \cellcolor[HTML]{18cde4} \textcolor[HTML]{000000}{201} & \cellcolor[HTML]{7c06ff} \textcolor[HTML]{f1f1f1}{8} & \cellcolor[HTML]{000000} \textcolor[HTML]{f1f1f1}{0} & \cellcolor[HTML]{000000} \textcolor[HTML]{f1f1f1}{0} & \cellcolor[HTML]{000000} \textcolor[HTML]{f1f1f1}{0} & \cellcolor[HTML]{000000} \textcolor[HTML]{f1f1f1}{0} \\
1 & \cellcolor[HTML]{7019ff} \textcolor[HTML]{f1f1f1}{22} & \cellcolor[HTML]{86ffb0} \textcolor[HTML]{000000}{344} & \cellcolor[HTML]{000000} \textcolor[HTML]{f1f1f1}{0} & \cellcolor[HTML]{000000} \textcolor[HTML]{f1f1f1}{0} & \cellcolor[HTML]{8000ff} \textcolor[HTML]{f1f1f1}{3} & \cellcolor[HTML]{7413ff} \textcolor[HTML]{f1f1f1}{17} \\
0 & \cellcolor[HTML]{000000} \textcolor[HTML]{f1f1f1}{0} & \cellcolor[HTML]{b4f295} \textcolor[HTML]{000000}{406} & \cellcolor[HTML]{000000} \textcolor[HTML]{f1f1f1}{0} & \cellcolor[HTML]{000000} \textcolor[HTML]{f1f1f1}{0} & \cellcolor[HTML]{000000} \textcolor[HTML]{f1f1f1}{0} & \cellcolor[HTML]{000000} \textcolor[HTML]{f1f1f1}{0} \\
7 & \cellcolor[HTML]{000000} \textcolor[HTML]{f1f1f1}{0} & \cellcolor[HTML]{4af2cf} \textcolor[HTML]{000000}{266} & \cellcolor[HTML]{000000} \textcolor[HTML]{f1f1f1}{0} & \cellcolor[HTML]{000000} \textcolor[HTML]{f1f1f1}{0} & \cellcolor[HTML]{8000ff} \textcolor[HTML]{f1f1f1}{2} & \cellcolor[HTML]{780dff} \textcolor[HTML]{f1f1f1}{13} \\
5 & \cellcolor[HTML]{000000} \textcolor[HTML]{f1f1f1}{0} & \cellcolor[HTML]{4659fb} \textcolor[HTML]{f1f1f1}{77} & \cellcolor[HTML]{5247fc} \textcolor[HTML]{f1f1f1}{63} & \cellcolor[HTML]{3c68f9} \textcolor[HTML]{f1f1f1}{92} & \cellcolor[HTML]{6826fe} \textcolor[HTML]{f1f1f1}{33} & \cellcolor[HTML]{7019ff} \textcolor[HTML]{f1f1f1}{22} \\
\end{tabular}

    }
    \hfill
    \parbox{35mm}{
        \centering
        \begin{tabular}{rrrrrrr}
Cl & WC & M & S & SR & R & SM \\
9 & \cellcolor[HTML]{14cae5} \textcolor[HTML]{000000}{195} & \cellcolor[HTML]{7c06ff} \textcolor[HTML]{f1f1f1}{8} & \cellcolor[HTML]{000000} \textcolor[HTML]{f1f1f1}{0} & \cellcolor[HTML]{000000} \textcolor[HTML]{f1f1f1}{0} & \cellcolor[HTML]{000000} \textcolor[HTML]{f1f1f1}{0} & \cellcolor[HTML]{000000} \textcolor[HTML]{f1f1f1}{0} \\
0 & \cellcolor[HTML]{7610ff} \textcolor[HTML]{f1f1f1}{15} & \cellcolor[HTML]{ffa95b} \textcolor[HTML]{000000}{514} & \cellcolor[HTML]{000000} \textcolor[HTML]{f1f1f1}{0} & \cellcolor[HTML]{000000} \textcolor[HTML]{f1f1f1}{0} & \cellcolor[HTML]{7c06ff} \textcolor[HTML]{f1f1f1}{7} & \cellcolor[HTML]{7019ff} \textcolor[HTML]{f1f1f1}{24} \\
4 & \cellcolor[HTML]{7e03ff} \textcolor[HTML]{f1f1f1}{4} & \cellcolor[HTML]{5df9c7} \textcolor[HTML]{000000}{291} & \cellcolor[HTML]{000000} \textcolor[HTML]{f1f1f1}{0} & \cellcolor[HTML]{000000} \textcolor[HTML]{f1f1f1}{0} & \cellcolor[HTML]{000000} \textcolor[HTML]{f1f1f1}{0} & \cellcolor[HTML]{8000ff} \textcolor[HTML]{f1f1f1}{2} \\
7 & \cellcolor[HTML]{000000} \textcolor[HTML]{f1f1f1}{0} & \cellcolor[HTML]{37e6d8} \textcolor[HTML]{000000}{240} & \cellcolor[HTML]{000000} \textcolor[HTML]{f1f1f1}{0} & \cellcolor[HTML]{000000} \textcolor[HTML]{f1f1f1}{0} & \cellcolor[HTML]{000000} \textcolor[HTML]{f1f1f1}{0} & \cellcolor[HTML]{000000} \textcolor[HTML]{f1f1f1}{0} \\
2 & \cellcolor[HTML]{000000} \textcolor[HTML]{f1f1f1}{0} & \cellcolor[HTML]{4e4dfc} \textcolor[HTML]{f1f1f1}{68} & \cellcolor[HTML]{5c38fd} \textcolor[HTML]{f1f1f1}{49} & \cellcolor[HTML]{1c93f3} \textcolor[HTML]{f1f1f1}{133} & \cellcolor[HTML]{5c38fd} \textcolor[HTML]{f1f1f1}{49} & \cellcolor[HTML]{7413ff} \textcolor[HTML]{f1f1f1}{17} \\
\end{tabular}

    }
    \label{tab:cluster_class_matrices_simclr_3000}
\end{table*}

\addtolength{\tabcolsep}{+0.17em}
\begin{table*}
    \caption{
        Cluster vs. ground truth class matrices for the same 5 repeat runs of remote summaries as in table \ref{tab:cluster_class_matrices_simclr_3000} but only using the clustering of the 50 representative images (5 per cluster) that are selected for transmission as compressed bitmaps, along with their metadata.
    }
    \parbox{35mm}{
        \centering
        \begin{tabular}{rrrrrrr}
Cl & WC & M & S & SR & R & SM \\
9 & \cellcolor[HTML]{ff0000} \textcolor[HTML]{f1f1f1}{5} & \cellcolor[HTML]{000000} \textcolor[HTML]{f1f1f1}{0} & \cellcolor[HTML]{000000} \textcolor[HTML]{f1f1f1}{0} & \cellcolor[HTML]{000000} \textcolor[HTML]{f1f1f1}{0} & \cellcolor[HTML]{000000} \textcolor[HTML]{f1f1f1}{0} & \cellcolor[HTML]{000000} \textcolor[HTML]{f1f1f1}{0} \\
0 & \cellcolor[HTML]{000000} \textcolor[HTML]{f1f1f1}{0} & \cellcolor[HTML]{ff0000} \textcolor[HTML]{f1f1f1}{5} & \cellcolor[HTML]{000000} \textcolor[HTML]{f1f1f1}{0} & \cellcolor[HTML]{000000} \textcolor[HTML]{f1f1f1}{0} & \cellcolor[HTML]{000000} \textcolor[HTML]{f1f1f1}{0} & \cellcolor[HTML]{000000} \textcolor[HTML]{f1f1f1}{0} \\
1 & \cellcolor[HTML]{000000} \textcolor[HTML]{f1f1f1}{0} & \cellcolor[HTML]{ff0000} \textcolor[HTML]{f1f1f1}{5} & \cellcolor[HTML]{000000} \textcolor[HTML]{f1f1f1}{0} & \cellcolor[HTML]{000000} \textcolor[HTML]{f1f1f1}{0} & \cellcolor[HTML]{000000} \textcolor[HTML]{f1f1f1}{0} & \cellcolor[HTML]{000000} \textcolor[HTML]{f1f1f1}{0} \\
5 & \cellcolor[HTML]{000000} \textcolor[HTML]{f1f1f1}{0} & \cellcolor[HTML]{ff0000} \textcolor[HTML]{f1f1f1}{5} & \cellcolor[HTML]{000000} \textcolor[HTML]{f1f1f1}{0} & \cellcolor[HTML]{000000} \textcolor[HTML]{f1f1f1}{0} & \cellcolor[HTML]{000000} \textcolor[HTML]{f1f1f1}{0} & \cellcolor[HTML]{000000} \textcolor[HTML]{f1f1f1}{0} \\
3 & \cellcolor[HTML]{000000} \textcolor[HTML]{f1f1f1}{0} & \cellcolor[HTML]{8000ff} \textcolor[HTML]{f1f1f1}{1} & \cellcolor[HTML]{000000} \textcolor[HTML]{f1f1f1}{0} & \cellcolor[HTML]{8000ff} \textcolor[HTML]{f1f1f1}{1} & \cellcolor[HTML]{80ffb4} \textcolor[HTML]{000000}{3} & \cellcolor[HTML]{000000} \textcolor[HTML]{f1f1f1}{0} \\
\end{tabular}

    }
    \hfill
    \parbox{35mm}{
        \centering
        \begin{tabular}{rrrrrrr}
Cl & WC & M & S & SR & R & SM \\
9 & \cellcolor[HTML]{ff0000} \textcolor[HTML]{f1f1f1}{5} & \cellcolor[HTML]{000000} \textcolor[HTML]{f1f1f1}{0} & \cellcolor[HTML]{000000} \textcolor[HTML]{f1f1f1}{0} & \cellcolor[HTML]{000000} \textcolor[HTML]{f1f1f1}{0} & \cellcolor[HTML]{000000} \textcolor[HTML]{f1f1f1}{0} & \cellcolor[HTML]{000000} \textcolor[HTML]{f1f1f1}{0} \\
8 & \cellcolor[HTML]{8000ff} \textcolor[HTML]{f1f1f1}{1} & \cellcolor[HTML]{ffb360} \textcolor[HTML]{000000}{4} & \cellcolor[HTML]{000000} \textcolor[HTML]{f1f1f1}{0} & \cellcolor[HTML]{000000} \textcolor[HTML]{f1f1f1}{0} & \cellcolor[HTML]{000000} \textcolor[HTML]{f1f1f1}{0} & \cellcolor[HTML]{000000} \textcolor[HTML]{f1f1f1}{0} \\
2 & \cellcolor[HTML]{000000} \textcolor[HTML]{f1f1f1}{0} & \cellcolor[HTML]{ff0000} \textcolor[HTML]{f1f1f1}{5} & \cellcolor[HTML]{000000} \textcolor[HTML]{f1f1f1}{0} & \cellcolor[HTML]{000000} \textcolor[HTML]{f1f1f1}{0} & \cellcolor[HTML]{000000} \textcolor[HTML]{f1f1f1}{0} & \cellcolor[HTML]{000000} \textcolor[HTML]{f1f1f1}{0} \\
5 & \cellcolor[HTML]{000000} \textcolor[HTML]{f1f1f1}{0} & \cellcolor[HTML]{ff0000} \textcolor[HTML]{f1f1f1}{5} & \cellcolor[HTML]{000000} \textcolor[HTML]{f1f1f1}{0} & \cellcolor[HTML]{000000} \textcolor[HTML]{f1f1f1}{0} & \cellcolor[HTML]{000000} \textcolor[HTML]{f1f1f1}{0} & \cellcolor[HTML]{000000} \textcolor[HTML]{f1f1f1}{0} \\
3 & \cellcolor[HTML]{000000} \textcolor[HTML]{f1f1f1}{0} & \cellcolor[HTML]{ffb360} \textcolor[HTML]{000000}{4} & \cellcolor[HTML]{000000} \textcolor[HTML]{f1f1f1}{0} & \cellcolor[HTML]{000000} \textcolor[HTML]{f1f1f1}{0} & \cellcolor[HTML]{000000} \textcolor[HTML]{f1f1f1}{0} & \cellcolor[HTML]{8000ff} \textcolor[HTML]{f1f1f1}{1} \\
\end{tabular}

    }
    \hfill
    \parbox{35mm}{
        \centering
        \begin{tabular}{rrrrrrr}
Cl & WC & M & S & SR & R & SM \\
6 & \cellcolor[HTML]{ff0000} \textcolor[HTML]{f1f1f1}{5} & \cellcolor[HTML]{000000} \textcolor[HTML]{f1f1f1}{0} & \cellcolor[HTML]{000000} \textcolor[HTML]{f1f1f1}{0} & \cellcolor[HTML]{000000} \textcolor[HTML]{f1f1f1}{0} & \cellcolor[HTML]{000000} \textcolor[HTML]{f1f1f1}{0} & \cellcolor[HTML]{000000} \textcolor[HTML]{f1f1f1}{0} \\
0 & \cellcolor[HTML]{000000} \textcolor[HTML]{f1f1f1}{0} & \cellcolor[HTML]{ff0000} \textcolor[HTML]{f1f1f1}{5} & \cellcolor[HTML]{000000} \textcolor[HTML]{f1f1f1}{0} & \cellcolor[HTML]{000000} \textcolor[HTML]{f1f1f1}{0} & \cellcolor[HTML]{000000} \textcolor[HTML]{f1f1f1}{0} & \cellcolor[HTML]{000000} \textcolor[HTML]{f1f1f1}{0} \\
1 & \cellcolor[HTML]{000000} \textcolor[HTML]{f1f1f1}{0} & \cellcolor[HTML]{ff0000} \textcolor[HTML]{f1f1f1}{5} & \cellcolor[HTML]{000000} \textcolor[HTML]{f1f1f1}{0} & \cellcolor[HTML]{000000} \textcolor[HTML]{f1f1f1}{0} & \cellcolor[HTML]{000000} \textcolor[HTML]{f1f1f1}{0} & \cellcolor[HTML]{000000} \textcolor[HTML]{f1f1f1}{0} \\
8 & \cellcolor[HTML]{000000} \textcolor[HTML]{f1f1f1}{0} & \cellcolor[HTML]{80ffb4} \textcolor[HTML]{000000}{3} & \cellcolor[HTML]{8000ff} \textcolor[HTML]{f1f1f1}{1} & \cellcolor[HTML]{000000} \textcolor[HTML]{f1f1f1}{0} & \cellcolor[HTML]{000000} \textcolor[HTML]{f1f1f1}{0} & \cellcolor[HTML]{8000ff} \textcolor[HTML]{f1f1f1}{1} \\
3 & \cellcolor[HTML]{000000} \textcolor[HTML]{f1f1f1}{0} & \cellcolor[HTML]{000000} \textcolor[HTML]{f1f1f1}{0} & \cellcolor[HTML]{ff0000} \textcolor[HTML]{f1f1f1}{5} & \cellcolor[HTML]{000000} \textcolor[HTML]{f1f1f1}{0} & \cellcolor[HTML]{000000} \textcolor[HTML]{f1f1f1}{0} & \cellcolor[HTML]{000000} \textcolor[HTML]{f1f1f1}{0} \\
\end{tabular}

    }
    \hfill
    \parbox{35mm}{
        \centering
        \begin{tabular}{rrrrrrr}
Cl & WC & M & S & SR & R & SM \\
9 & \cellcolor[HTML]{ff0000} \textcolor[HTML]{f1f1f1}{5} & \cellcolor[HTML]{000000} \textcolor[HTML]{f1f1f1}{0} & \cellcolor[HTML]{000000} \textcolor[HTML]{f1f1f1}{0} & \cellcolor[HTML]{000000} \textcolor[HTML]{f1f1f1}{0} & \cellcolor[HTML]{000000} \textcolor[HTML]{f1f1f1}{0} & \cellcolor[HTML]{000000} \textcolor[HTML]{f1f1f1}{0} \\
0 & \cellcolor[HTML]{000000} \textcolor[HTML]{f1f1f1}{0} & \cellcolor[HTML]{ff0000} \textcolor[HTML]{f1f1f1}{5} & \cellcolor[HTML]{000000} \textcolor[HTML]{f1f1f1}{0} & \cellcolor[HTML]{000000} \textcolor[HTML]{f1f1f1}{0} & \cellcolor[HTML]{000000} \textcolor[HTML]{f1f1f1}{0} & \cellcolor[HTML]{000000} \textcolor[HTML]{f1f1f1}{0} \\
1 & \cellcolor[HTML]{000000} \textcolor[HTML]{f1f1f1}{0} & \cellcolor[HTML]{ff0000} \textcolor[HTML]{f1f1f1}{5} & \cellcolor[HTML]{000000} \textcolor[HTML]{f1f1f1}{0} & \cellcolor[HTML]{000000} \textcolor[HTML]{f1f1f1}{0} & \cellcolor[HTML]{000000} \textcolor[HTML]{f1f1f1}{0} & \cellcolor[HTML]{000000} \textcolor[HTML]{f1f1f1}{0} \\
7 & \cellcolor[HTML]{000000} \textcolor[HTML]{f1f1f1}{0} & \cellcolor[HTML]{ff0000} \textcolor[HTML]{f1f1f1}{5} & \cellcolor[HTML]{000000} \textcolor[HTML]{f1f1f1}{0} & \cellcolor[HTML]{000000} \textcolor[HTML]{f1f1f1}{0} & \cellcolor[HTML]{000000} \textcolor[HTML]{f1f1f1}{0} & \cellcolor[HTML]{000000} \textcolor[HTML]{f1f1f1}{0} \\
5 & \cellcolor[HTML]{000000} \textcolor[HTML]{f1f1f1}{0} & \cellcolor[HTML]{00b5eb} \textcolor[HTML]{f1f1f1}{2} & \cellcolor[HTML]{000000} \textcolor[HTML]{f1f1f1}{0} & \cellcolor[HTML]{80ffb4} \textcolor[HTML]{000000}{3} & \cellcolor[HTML]{000000} \textcolor[HTML]{f1f1f1}{0} & \cellcolor[HTML]{000000} \textcolor[HTML]{f1f1f1}{0} \\
\end{tabular}

    }
    \hfill
    \parbox{35mm}{
        \centering
        \begin{tabular}{rrrrrrr}
Cl & WC & M & S & SR & R & SM \\
9 & \cellcolor[HTML]{ff0000} \textcolor[HTML]{f1f1f1}{5} & \cellcolor[HTML]{000000} \textcolor[HTML]{f1f1f1}{0} & \cellcolor[HTML]{000000} \textcolor[HTML]{f1f1f1}{0} & \cellcolor[HTML]{000000} \textcolor[HTML]{f1f1f1}{0} & \cellcolor[HTML]{000000} \textcolor[HTML]{f1f1f1}{0} & \cellcolor[HTML]{000000} \textcolor[HTML]{f1f1f1}{0} \\
0 & \cellcolor[HTML]{000000} \textcolor[HTML]{f1f1f1}{0} & \cellcolor[HTML]{ff0000} \textcolor[HTML]{f1f1f1}{5} & \cellcolor[HTML]{000000} \textcolor[HTML]{f1f1f1}{0} & \cellcolor[HTML]{000000} \textcolor[HTML]{f1f1f1}{0} & \cellcolor[HTML]{000000} \textcolor[HTML]{f1f1f1}{0} & \cellcolor[HTML]{000000} \textcolor[HTML]{f1f1f1}{0} \\
4 & \cellcolor[HTML]{000000} \textcolor[HTML]{f1f1f1}{0} & \cellcolor[HTML]{ff0000} \textcolor[HTML]{f1f1f1}{5} & \cellcolor[HTML]{000000} \textcolor[HTML]{f1f1f1}{0} & \cellcolor[HTML]{000000} \textcolor[HTML]{f1f1f1}{0} & \cellcolor[HTML]{000000} \textcolor[HTML]{f1f1f1}{0} & \cellcolor[HTML]{000000} \textcolor[HTML]{f1f1f1}{0} \\
7 & \cellcolor[HTML]{000000} \textcolor[HTML]{f1f1f1}{0} & \cellcolor[HTML]{ff0000} \textcolor[HTML]{f1f1f1}{5} & \cellcolor[HTML]{000000} \textcolor[HTML]{f1f1f1}{0} & \cellcolor[HTML]{000000} \textcolor[HTML]{f1f1f1}{0} & \cellcolor[HTML]{000000} \textcolor[HTML]{f1f1f1}{0} & \cellcolor[HTML]{000000} \textcolor[HTML]{f1f1f1}{0} \\
1 & \cellcolor[HTML]{000000} \textcolor[HTML]{f1f1f1}{0} & \cellcolor[HTML]{8000ff} \textcolor[HTML]{f1f1f1}{1} & \cellcolor[HTML]{80ffb4} \textcolor[HTML]{000000}{3} & \cellcolor[HTML]{8000ff} \textcolor[HTML]{f1f1f1}{1} & \cellcolor[HTML]{000000} \textcolor[HTML]{f1f1f1}{0} & \cellcolor[HTML]{000000} \textcolor[HTML]{f1f1f1}{0} \\
\end{tabular}

    }
    \label{tab:cluster_class_matrices_simclr_50}
\end{table*}
\addtolength{\tabcolsep}{-0.17em}

\begin{table*}
    \caption{
        Cluster vs. ground truth class matrices for 5 repeat runs of remote summaries using 3000 random sampled images, encoded with the DINO model.
    }
    \parbox{35mm}{
        \centering
        \begin{tabular}{rrrrrrr}
Cl & WC & M & S & SR & R & SM \\
8 & \cellcolor[HTML]{18cde4} \textcolor[HTML]{000000}{202} & \cellcolor[HTML]{7e03ff} \textcolor[HTML]{f1f1f1}{4} & \cellcolor[HTML]{000000} \textcolor[HTML]{f1f1f1}{0} & \cellcolor[HTML]{000000} \textcolor[HTML]{f1f1f1}{0} & \cellcolor[HTML]{000000} \textcolor[HTML]{f1f1f1}{0} & \cellcolor[HTML]{000000} \textcolor[HTML]{f1f1f1}{0} \\
0 & \cellcolor[HTML]{000000} \textcolor[HTML]{f1f1f1}{0} & \cellcolor[HTML]{ff0000} \textcolor[HTML]{f1f1f1}{670} & \cellcolor[HTML]{000000} \textcolor[HTML]{f1f1f1}{0} & \cellcolor[HTML]{000000} \textcolor[HTML]{f1f1f1}{0} & \cellcolor[HTML]{8000ff} \textcolor[HTML]{f1f1f1}{2} & \cellcolor[HTML]{7c06ff} \textcolor[HTML]{f1f1f1}{7} \\
1 & \cellcolor[HTML]{000000} \textcolor[HTML]{f1f1f1}{0} & \cellcolor[HTML]{aef498} \textcolor[HTML]{000000}{398} & \cellcolor[HTML]{000000} \textcolor[HTML]{f1f1f1}{0} & \cellcolor[HTML]{000000} \textcolor[HTML]{f1f1f1}{0} & \cellcolor[HTML]{8000ff} \textcolor[HTML]{f1f1f1}{3} & \cellcolor[HTML]{7e03ff} \textcolor[HTML]{f1f1f1}{5} \\
9 & \cellcolor[HTML]{000000} \textcolor[HTML]{f1f1f1}{0} & \cellcolor[HTML]{504afc} \textcolor[HTML]{f1f1f1}{65} & \cellcolor[HTML]{7019ff} \textcolor[HTML]{f1f1f1}{23} & \cellcolor[HTML]{8000ff} \textcolor[HTML]{f1f1f1}{1} & \cellcolor[HTML]{8000ff} \textcolor[HTML]{f1f1f1}{2} & \cellcolor[HTML]{5c38fd} \textcolor[HTML]{f1f1f1}{49} \\
4 & \cellcolor[HTML]{000000} \textcolor[HTML]{f1f1f1}{0} & \cellcolor[HTML]{5a3bfd} \textcolor[HTML]{f1f1f1}{53} & \cellcolor[HTML]{0dc2e8} \textcolor[HTML]{000000}{185} & \cellcolor[HTML]{7a09ff} \textcolor[HTML]{f1f1f1}{11} & \cellcolor[HTML]{000000} \textcolor[HTML]{f1f1f1}{0} & \cellcolor[HTML]{7e03ff} \textcolor[HTML]{f1f1f1}{5} \\
\end{tabular}

    }
    \hfill
    \parbox{35mm}{
        \centering
        \begin{tabular}{rrrrrrr}
Cl & WC & M & S & SR & R & SM \\
5 & \cellcolor[HTML]{32e3da} \textcolor[HTML]{000000}{235} & \cellcolor[HTML]{7e03ff} \textcolor[HTML]{f1f1f1}{4} & \cellcolor[HTML]{000000} \textcolor[HTML]{f1f1f1}{0} & \cellcolor[HTML]{000000} \textcolor[HTML]{f1f1f1}{0} & \cellcolor[HTML]{000000} \textcolor[HTML]{f1f1f1}{0} & \cellcolor[HTML]{000000} \textcolor[HTML]{f1f1f1}{0} \\
0 & \cellcolor[HTML]{000000} \textcolor[HTML]{f1f1f1}{0} & \cellcolor[HTML]{ff1c0e} \textcolor[HTML]{f1f1f1}{646} & \cellcolor[HTML]{000000} \textcolor[HTML]{f1f1f1}{0} & \cellcolor[HTML]{000000} \textcolor[HTML]{f1f1f1}{0} & \cellcolor[HTML]{8000ff} \textcolor[HTML]{f1f1f1}{1} & \cellcolor[HTML]{7a09ff} \textcolor[HTML]{f1f1f1}{9} \\
1 & \cellcolor[HTML]{000000} \textcolor[HTML]{f1f1f1}{0} & \cellcolor[HTML]{a8f79c} \textcolor[HTML]{000000}{388} & \cellcolor[HTML]{000000} \textcolor[HTML]{f1f1f1}{0} & \cellcolor[HTML]{000000} \textcolor[HTML]{f1f1f1}{0} & \cellcolor[HTML]{7e03ff} \textcolor[HTML]{f1f1f1}{4} & \cellcolor[HTML]{8000ff} \textcolor[HTML]{f1f1f1}{1} \\
8 & \cellcolor[HTML]{000000} \textcolor[HTML]{f1f1f1}{0} & \cellcolor[HTML]{4e4dfc} \textcolor[HTML]{f1f1f1}{68} & \cellcolor[HTML]{7019ff} \textcolor[HTML]{f1f1f1}{23} & \cellcolor[HTML]{8000ff} \textcolor[HTML]{f1f1f1}{1} & \cellcolor[HTML]{7c06ff} \textcolor[HTML]{f1f1f1}{8} & \cellcolor[HTML]{5641fd} \textcolor[HTML]{f1f1f1}{56} \\
6 & \cellcolor[HTML]{000000} \textcolor[HTML]{f1f1f1}{0} & \cellcolor[HTML]{5641fd} \textcolor[HTML]{f1f1f1}{56} & \cellcolor[HTML]{04b0ed} \textcolor[HTML]{f1f1f1}{164} & \cellcolor[HTML]{7a09ff} \textcolor[HTML]{f1f1f1}{9} & \cellcolor[HTML]{000000} \textcolor[HTML]{f1f1f1}{0} & \cellcolor[HTML]{8000ff} \textcolor[HTML]{f1f1f1}{1} \\
\end{tabular}

    }
    \hfill
    \parbox{35mm}{
        \centering
        \begin{tabular}{rrrrrrr}
Cl & WC & M & S & SR & R & SM \\
7 & \cellcolor[HTML]{2ddedc} \textcolor[HTML]{000000}{227} & \cellcolor[HTML]{7e03ff} \textcolor[HTML]{f1f1f1}{5} & \cellcolor[HTML]{000000} \textcolor[HTML]{f1f1f1}{0} & \cellcolor[HTML]{000000} \textcolor[HTML]{f1f1f1}{0} & \cellcolor[HTML]{000000} \textcolor[HTML]{f1f1f1}{0} & \cellcolor[HTML]{000000} \textcolor[HTML]{f1f1f1}{0} \\
0 & \cellcolor[HTML]{000000} \textcolor[HTML]{f1f1f1}{0} & \cellcolor[HTML]{ff8444} \textcolor[HTML]{f1f1f1}{553} & \cellcolor[HTML]{000000} \textcolor[HTML]{f1f1f1}{0} & \cellcolor[HTML]{000000} \textcolor[HTML]{f1f1f1}{0} & \cellcolor[HTML]{8000ff} \textcolor[HTML]{f1f1f1}{1} & \cellcolor[HTML]{780dff} \textcolor[HTML]{f1f1f1}{13} \\
2 & \cellcolor[HTML]{000000} \textcolor[HTML]{f1f1f1}{0} & \cellcolor[HTML]{a2f9a0} \textcolor[HTML]{000000}{380} & \cellcolor[HTML]{000000} \textcolor[HTML]{f1f1f1}{0} & \cellcolor[HTML]{000000} \textcolor[HTML]{f1f1f1}{0} & \cellcolor[HTML]{8000ff} \textcolor[HTML]{f1f1f1}{1} & \cellcolor[HTML]{8000ff} \textcolor[HTML]{f1f1f1}{3} \\
9 & \cellcolor[HTML]{000000} \textcolor[HTML]{f1f1f1}{0} & \cellcolor[HTML]{2489f5} \textcolor[HTML]{f1f1f1}{122} & \cellcolor[HTML]{000000} \textcolor[HTML]{f1f1f1}{0} & \cellcolor[HTML]{000000} \textcolor[HTML]{f1f1f1}{0} & \cellcolor[HTML]{000000} \textcolor[HTML]{f1f1f1}{0} & \cellcolor[HTML]{000000} \textcolor[HTML]{f1f1f1}{0} \\
8 & \cellcolor[HTML]{000000} \textcolor[HTML]{f1f1f1}{0} & \cellcolor[HTML]{4c50fc} \textcolor[HTML]{f1f1f1}{69} & \cellcolor[HTML]{6e1cff} \textcolor[HTML]{f1f1f1}{27} & \cellcolor[HTML]{8000ff} \textcolor[HTML]{f1f1f1}{3} & \cellcolor[HTML]{7e03ff} \textcolor[HTML]{f1f1f1}{5} & \cellcolor[HTML]{5e35fe} \textcolor[HTML]{f1f1f1}{48} \\
\end{tabular}

    }
    \hfill
    \parbox{35mm}{
        \centering
        \begin{tabular}{rrrrrrr}
Cl & WC & M & S & SR & R & SM \\
7 & \cellcolor[HTML]{22d6e0} \textcolor[HTML]{000000}{213} & \cellcolor[HTML]{8000ff} \textcolor[HTML]{f1f1f1}{2} & \cellcolor[HTML]{000000} \textcolor[HTML]{f1f1f1}{0} & \cellcolor[HTML]{000000} \textcolor[HTML]{f1f1f1}{0} & \cellcolor[HTML]{000000} \textcolor[HTML]{f1f1f1}{0} & \cellcolor[HTML]{000000} \textcolor[HTML]{f1f1f1}{0} \\
0 & \cellcolor[HTML]{000000} \textcolor[HTML]{f1f1f1}{0} & \cellcolor[HTML]{ff0000} \textcolor[HTML]{f1f1f1}{671} & \cellcolor[HTML]{000000} \textcolor[HTML]{f1f1f1}{0} & \cellcolor[HTML]{000000} \textcolor[HTML]{f1f1f1}{0} & \cellcolor[HTML]{8000ff} \textcolor[HTML]{f1f1f1}{2} & \cellcolor[HTML]{7a09ff} \textcolor[HTML]{f1f1f1}{10} \\
1 & \cellcolor[HTML]{000000} \textcolor[HTML]{f1f1f1}{0} & \cellcolor[HTML]{d4dd80} \textcolor[HTML]{000000}{446} & \cellcolor[HTML]{7e03ff} \textcolor[HTML]{f1f1f1}{5} & \cellcolor[HTML]{000000} \textcolor[HTML]{f1f1f1}{0} & \cellcolor[HTML]{7e03ff} \textcolor[HTML]{f1f1f1}{5} & \cellcolor[HTML]{642cfe} \textcolor[HTML]{f1f1f1}{39} \\
3 & \cellcolor[HTML]{000000} \textcolor[HTML]{f1f1f1}{0} & \cellcolor[HTML]{5444fd} \textcolor[HTML]{f1f1f1}{61} & \cellcolor[HTML]{10c6e6} \textcolor[HTML]{000000}{190} & \cellcolor[HTML]{7a09ff} \textcolor[HTML]{f1f1f1}{10} & \cellcolor[HTML]{000000} \textcolor[HTML]{f1f1f1}{0} & \cellcolor[HTML]{7e03ff} \textcolor[HTML]{f1f1f1}{4} \\
4 & \cellcolor[HTML]{000000} \textcolor[HTML]{f1f1f1}{0} & \cellcolor[HTML]{8000ff} \textcolor[HTML]{f1f1f1}{1} & \cellcolor[HTML]{30e1da} \textcolor[HTML]{000000}{231} & \cellcolor[HTML]{8000ff} \textcolor[HTML]{f1f1f1}{3} & \cellcolor[HTML]{000000} \textcolor[HTML]{f1f1f1}{0} & \cellcolor[HTML]{7413ff} \textcolor[HTML]{f1f1f1}{17} \\
\end{tabular}

    }
    \hfill
    \parbox{35mm}{
        \centering
        \begin{tabular}{rrrrrrr}
Cl & WC & M & S & SR & R & SM \\
6 & \cellcolor[HTML]{30e1da} \textcolor[HTML]{000000}{233} & \cellcolor[HTML]{7e03ff} \textcolor[HTML]{f1f1f1}{5} & \cellcolor[HTML]{000000} \textcolor[HTML]{f1f1f1}{0} & \cellcolor[HTML]{000000} \textcolor[HTML]{f1f1f1}{0} & \cellcolor[HTML]{000000} \textcolor[HTML]{f1f1f1}{0} & \cellcolor[HTML]{000000} \textcolor[HTML]{f1f1f1}{0} \\
0 & \cellcolor[HTML]{000000} \textcolor[HTML]{f1f1f1}{0} & \cellcolor[HTML]{ff0603} \textcolor[HTML]{f1f1f1}{664} & \cellcolor[HTML]{000000} \textcolor[HTML]{f1f1f1}{0} & \cellcolor[HTML]{000000} \textcolor[HTML]{f1f1f1}{0} & \cellcolor[HTML]{8000ff} \textcolor[HTML]{f1f1f1}{2} & \cellcolor[HTML]{7a09ff} \textcolor[HTML]{f1f1f1}{9} \\
1 & \cellcolor[HTML]{000000} \textcolor[HTML]{f1f1f1}{0} & \cellcolor[HTML]{abf69b} \textcolor[HTML]{000000}{391} & \cellcolor[HTML]{000000} \textcolor[HTML]{f1f1f1}{0} & \cellcolor[HTML]{000000} \textcolor[HTML]{f1f1f1}{0} & \cellcolor[HTML]{8000ff} \textcolor[HTML]{f1f1f1}{3} & \cellcolor[HTML]{8000ff} \textcolor[HTML]{f1f1f1}{3} \\
8 & \cellcolor[HTML]{000000} \textcolor[HTML]{f1f1f1}{0} & \cellcolor[HTML]{4e4dfc} \textcolor[HTML]{f1f1f1}{68} & \cellcolor[HTML]{6e1cff} \textcolor[HTML]{f1f1f1}{25} & \cellcolor[HTML]{7e03ff} \textcolor[HTML]{f1f1f1}{4} & \cellcolor[HTML]{7c06ff} \textcolor[HTML]{f1f1f1}{7} & \cellcolor[HTML]{5a3bfd} \textcolor[HTML]{f1f1f1}{52} \\
5 & \cellcolor[HTML]{000000} \textcolor[HTML]{f1f1f1}{0} & \cellcolor[HTML]{5a3bfd} \textcolor[HTML]{f1f1f1}{52} & \cellcolor[HTML]{04b9ea} \textcolor[HTML]{f1f1f1}{176} & \cellcolor[HTML]{7610ff} \textcolor[HTML]{f1f1f1}{15} & \cellcolor[HTML]{000000} \textcolor[HTML]{f1f1f1}{0} & \cellcolor[HTML]{8000ff} \textcolor[HTML]{f1f1f1}{3} \\
\end{tabular}

    }
    \label{tab:cluster_class_matrices_dino_3000}
\end{table*}

\addtolength{\tabcolsep}{+0.17em}
\begin{table*}
    \caption{
        Cluster vs. ground truth class matrices for the same 5 repeat runs of remote summaries as in table \ref{tab:cluster_class_matrices_dino_3000} but only using the clustering of the 50 representative images (5 per cluster) that are selected for transmission as compressed bitmaps, along with their metadata.
    }
    \parbox{35mm}{
        \centering
        \begin{tabular}{rrrrrrr}
Cl & WC & M & S & SR & R & SM \\
8 & \cellcolor[HTML]{ff0000} \textcolor[HTML]{f1f1f1}{5} & \cellcolor[HTML]{000000} \textcolor[HTML]{f1f1f1}{0} & \cellcolor[HTML]{000000} \textcolor[HTML]{f1f1f1}{0} & \cellcolor[HTML]{000000} \textcolor[HTML]{f1f1f1}{0} & \cellcolor[HTML]{000000} \textcolor[HTML]{f1f1f1}{0} & \cellcolor[HTML]{000000} \textcolor[HTML]{f1f1f1}{0} \\
0 & \cellcolor[HTML]{000000} \textcolor[HTML]{f1f1f1}{0} & \cellcolor[HTML]{ff0000} \textcolor[HTML]{f1f1f1}{5} & \cellcolor[HTML]{000000} \textcolor[HTML]{f1f1f1}{0} & \cellcolor[HTML]{000000} \textcolor[HTML]{f1f1f1}{0} & \cellcolor[HTML]{000000} \textcolor[HTML]{f1f1f1}{0} & \cellcolor[HTML]{000000} \textcolor[HTML]{f1f1f1}{0} \\
1 & \cellcolor[HTML]{000000} \textcolor[HTML]{f1f1f1}{0} & \cellcolor[HTML]{ff0000} \textcolor[HTML]{f1f1f1}{5} & \cellcolor[HTML]{000000} \textcolor[HTML]{f1f1f1}{0} & \cellcolor[HTML]{000000} \textcolor[HTML]{f1f1f1}{0} & \cellcolor[HTML]{000000} \textcolor[HTML]{f1f1f1}{0} & \cellcolor[HTML]{000000} \textcolor[HTML]{f1f1f1}{0} \\
9 & \cellcolor[HTML]{000000} \textcolor[HTML]{f1f1f1}{0} & \cellcolor[HTML]{8000ff} \textcolor[HTML]{f1f1f1}{1} & \cellcolor[HTML]{8000ff} \textcolor[HTML]{f1f1f1}{1} & \cellcolor[HTML]{000000} \textcolor[HTML]{f1f1f1}{0} & \cellcolor[HTML]{000000} \textcolor[HTML]{f1f1f1}{0} & \cellcolor[HTML]{80ffb4} \textcolor[HTML]{000000}{3} \\
2 & \cellcolor[HTML]{000000} \textcolor[HTML]{f1f1f1}{0} & \cellcolor[HTML]{000000} \textcolor[HTML]{f1f1f1}{0} & \cellcolor[HTML]{ff0000} \textcolor[HTML]{f1f1f1}{5} & \cellcolor[HTML]{000000} \textcolor[HTML]{f1f1f1}{0} & \cellcolor[HTML]{000000} \textcolor[HTML]{f1f1f1}{0} & \cellcolor[HTML]{000000} \textcolor[HTML]{f1f1f1}{0} \\
\end{tabular}

    }
    \hfill
    \parbox{35mm}{
        \centering
        \begin{tabular}{rrrrrrr}
Cl & WC & M & S & SR & R & SM \\
5 & \cellcolor[HTML]{ff0000} \textcolor[HTML]{f1f1f1}{5} & \cellcolor[HTML]{000000} \textcolor[HTML]{f1f1f1}{0} & \cellcolor[HTML]{000000} \textcolor[HTML]{f1f1f1}{0} & \cellcolor[HTML]{000000} \textcolor[HTML]{f1f1f1}{0} & \cellcolor[HTML]{000000} \textcolor[HTML]{f1f1f1}{0} & \cellcolor[HTML]{000000} \textcolor[HTML]{f1f1f1}{0} \\
0 & \cellcolor[HTML]{000000} \textcolor[HTML]{f1f1f1}{0} & \cellcolor[HTML]{ff0000} \textcolor[HTML]{f1f1f1}{5} & \cellcolor[HTML]{000000} \textcolor[HTML]{f1f1f1}{0} & \cellcolor[HTML]{000000} \textcolor[HTML]{f1f1f1}{0} & \cellcolor[HTML]{000000} \textcolor[HTML]{f1f1f1}{0} & \cellcolor[HTML]{000000} \textcolor[HTML]{f1f1f1}{0} \\
1 & \cellcolor[HTML]{000000} \textcolor[HTML]{f1f1f1}{0} & \cellcolor[HTML]{ff0000} \textcolor[HTML]{f1f1f1}{5} & \cellcolor[HTML]{000000} \textcolor[HTML]{f1f1f1}{0} & \cellcolor[HTML]{000000} \textcolor[HTML]{f1f1f1}{0} & \cellcolor[HTML]{000000} \textcolor[HTML]{f1f1f1}{0} & \cellcolor[HTML]{000000} \textcolor[HTML]{f1f1f1}{0} \\
6 & \cellcolor[HTML]{000000} \textcolor[HTML]{f1f1f1}{0} & \cellcolor[HTML]{00b5eb} \textcolor[HTML]{f1f1f1}{2} & \cellcolor[HTML]{80ffb4} \textcolor[HTML]{000000}{3} & \cellcolor[HTML]{000000} \textcolor[HTML]{f1f1f1}{0} & \cellcolor[HTML]{000000} \textcolor[HTML]{f1f1f1}{0} & \cellcolor[HTML]{000000} \textcolor[HTML]{f1f1f1}{0} \\
8 & \cellcolor[HTML]{000000} \textcolor[HTML]{f1f1f1}{0} & \cellcolor[HTML]{00b5eb} \textcolor[HTML]{f1f1f1}{2} & \cellcolor[HTML]{00b5eb} \textcolor[HTML]{f1f1f1}{2} & \cellcolor[HTML]{000000} \textcolor[HTML]{f1f1f1}{0} & \cellcolor[HTML]{000000} \textcolor[HTML]{f1f1f1}{0} & \cellcolor[HTML]{8000ff} \textcolor[HTML]{f1f1f1}{1} \\
\end{tabular}

    }
    \hfill
    \parbox{35mm}{
        \centering
        \begin{tabular}{rrrrrrr}
Cl & WC & M & S & SR & R & SM \\
7 & \cellcolor[HTML]{ff0000} \textcolor[HTML]{f1f1f1}{5} & \cellcolor[HTML]{000000} \textcolor[HTML]{f1f1f1}{0} & \cellcolor[HTML]{000000} \textcolor[HTML]{f1f1f1}{0} & \cellcolor[HTML]{000000} \textcolor[HTML]{f1f1f1}{0} & \cellcolor[HTML]{000000} \textcolor[HTML]{f1f1f1}{0} & \cellcolor[HTML]{000000} \textcolor[HTML]{f1f1f1}{0} \\
0 & \cellcolor[HTML]{000000} \textcolor[HTML]{f1f1f1}{0} & \cellcolor[HTML]{ff0000} \textcolor[HTML]{f1f1f1}{5} & \cellcolor[HTML]{000000} \textcolor[HTML]{f1f1f1}{0} & \cellcolor[HTML]{000000} \textcolor[HTML]{f1f1f1}{0} & \cellcolor[HTML]{000000} \textcolor[HTML]{f1f1f1}{0} & \cellcolor[HTML]{000000} \textcolor[HTML]{f1f1f1}{0} \\
2 & \cellcolor[HTML]{000000} \textcolor[HTML]{f1f1f1}{0} & \cellcolor[HTML]{ff0000} \textcolor[HTML]{f1f1f1}{5} & \cellcolor[HTML]{000000} \textcolor[HTML]{f1f1f1}{0} & \cellcolor[HTML]{000000} \textcolor[HTML]{f1f1f1}{0} & \cellcolor[HTML]{000000} \textcolor[HTML]{f1f1f1}{0} & \cellcolor[HTML]{000000} \textcolor[HTML]{f1f1f1}{0} \\
9 & \cellcolor[HTML]{000000} \textcolor[HTML]{f1f1f1}{0} & \cellcolor[HTML]{ff0000} \textcolor[HTML]{f1f1f1}{5} & \cellcolor[HTML]{000000} \textcolor[HTML]{f1f1f1}{0} & \cellcolor[HTML]{000000} \textcolor[HTML]{f1f1f1}{0} & \cellcolor[HTML]{000000} \textcolor[HTML]{f1f1f1}{0} & \cellcolor[HTML]{000000} \textcolor[HTML]{f1f1f1}{0} \\
6 & \cellcolor[HTML]{000000} \textcolor[HTML]{f1f1f1}{0} & \cellcolor[HTML]{00b5eb} \textcolor[HTML]{f1f1f1}{2} & \cellcolor[HTML]{80ffb4} \textcolor[HTML]{000000}{3} & \cellcolor[HTML]{000000} \textcolor[HTML]{f1f1f1}{0} & \cellcolor[HTML]{000000} \textcolor[HTML]{f1f1f1}{0} & \cellcolor[HTML]{000000} \textcolor[HTML]{f1f1f1}{0} \\
\end{tabular}

    }
    \hfill
    \parbox{35mm}{
        \centering
        \begin{tabular}{rrrrrrr}
Cl & WC & M & S & SR & R & SM \\
7 & \cellcolor[HTML]{ff0000} \textcolor[HTML]{f1f1f1}{5} & \cellcolor[HTML]{000000} \textcolor[HTML]{f1f1f1}{0} & \cellcolor[HTML]{000000} \textcolor[HTML]{f1f1f1}{0} & \cellcolor[HTML]{000000} \textcolor[HTML]{f1f1f1}{0} & \cellcolor[HTML]{000000} \textcolor[HTML]{f1f1f1}{0} & \cellcolor[HTML]{000000} \textcolor[HTML]{f1f1f1}{0} \\
0 & \cellcolor[HTML]{000000} \textcolor[HTML]{f1f1f1}{0} & \cellcolor[HTML]{ff0000} \textcolor[HTML]{f1f1f1}{5} & \cellcolor[HTML]{000000} \textcolor[HTML]{f1f1f1}{0} & \cellcolor[HTML]{000000} \textcolor[HTML]{f1f1f1}{0} & \cellcolor[HTML]{000000} \textcolor[HTML]{f1f1f1}{0} & \cellcolor[HTML]{000000} \textcolor[HTML]{f1f1f1}{0} \\
1 & \cellcolor[HTML]{000000} \textcolor[HTML]{f1f1f1}{0} & \cellcolor[HTML]{ff0000} \textcolor[HTML]{f1f1f1}{5} & \cellcolor[HTML]{000000} \textcolor[HTML]{f1f1f1}{0} & \cellcolor[HTML]{000000} \textcolor[HTML]{f1f1f1}{0} & \cellcolor[HTML]{000000} \textcolor[HTML]{f1f1f1}{0} & \cellcolor[HTML]{000000} \textcolor[HTML]{f1f1f1}{0} \\
3 & \cellcolor[HTML]{000000} \textcolor[HTML]{f1f1f1}{0} & \cellcolor[HTML]{00b5eb} \textcolor[HTML]{f1f1f1}{2} & \cellcolor[HTML]{80ffb4} \textcolor[HTML]{000000}{3} & \cellcolor[HTML]{000000} \textcolor[HTML]{f1f1f1}{0} & \cellcolor[HTML]{000000} \textcolor[HTML]{f1f1f1}{0} & \cellcolor[HTML]{000000} \textcolor[HTML]{f1f1f1}{0} \\
2 & \cellcolor[HTML]{000000} \textcolor[HTML]{f1f1f1}{0} & \cellcolor[HTML]{000000} \textcolor[HTML]{f1f1f1}{0} & \cellcolor[HTML]{ff0000} \textcolor[HTML]{f1f1f1}{5} & \cellcolor[HTML]{000000} \textcolor[HTML]{f1f1f1}{0} & \cellcolor[HTML]{000000} \textcolor[HTML]{f1f1f1}{0} & \cellcolor[HTML]{000000} \textcolor[HTML]{f1f1f1}{0} \\
\end{tabular}

    }
    \hfill
    \parbox{35mm}{
        \centering
        \begin{tabular}{rrrrrrr}
Cl & WC & M & S & SR & R & SM \\
6 & \cellcolor[HTML]{ff0000} \textcolor[HTML]{f1f1f1}{5} & \cellcolor[HTML]{000000} \textcolor[HTML]{f1f1f1}{0} & \cellcolor[HTML]{000000} \textcolor[HTML]{f1f1f1}{0} & \cellcolor[HTML]{000000} \textcolor[HTML]{f1f1f1}{0} & \cellcolor[HTML]{000000} \textcolor[HTML]{f1f1f1}{0} & \cellcolor[HTML]{000000} \textcolor[HTML]{f1f1f1}{0} \\
0 & \cellcolor[HTML]{000000} \textcolor[HTML]{f1f1f1}{0} & \cellcolor[HTML]{ff0000} \textcolor[HTML]{f1f1f1}{5} & \cellcolor[HTML]{000000} \textcolor[HTML]{f1f1f1}{0} & \cellcolor[HTML]{000000} \textcolor[HTML]{f1f1f1}{0} & \cellcolor[HTML]{000000} \textcolor[HTML]{f1f1f1}{0} & \cellcolor[HTML]{000000} \textcolor[HTML]{f1f1f1}{0} \\
1 & \cellcolor[HTML]{000000} \textcolor[HTML]{f1f1f1}{0} & \cellcolor[HTML]{ff0000} \textcolor[HTML]{f1f1f1}{5} & \cellcolor[HTML]{000000} \textcolor[HTML]{f1f1f1}{0} & \cellcolor[HTML]{000000} \textcolor[HTML]{f1f1f1}{0} & \cellcolor[HTML]{000000} \textcolor[HTML]{f1f1f1}{0} & \cellcolor[HTML]{000000} \textcolor[HTML]{f1f1f1}{0} \\
8 & \cellcolor[HTML]{000000} \textcolor[HTML]{f1f1f1}{0} & \cellcolor[HTML]{00b5eb} \textcolor[HTML]{f1f1f1}{2} & \cellcolor[HTML]{8000ff} \textcolor[HTML]{f1f1f1}{1} & \cellcolor[HTML]{000000} \textcolor[HTML]{f1f1f1}{0} & \cellcolor[HTML]{000000} \textcolor[HTML]{f1f1f1}{0} & \cellcolor[HTML]{00b5eb} \textcolor[HTML]{f1f1f1}{2} \\
5 & \cellcolor[HTML]{000000} \textcolor[HTML]{f1f1f1}{0} & \cellcolor[HTML]{8000ff} \textcolor[HTML]{f1f1f1}{1} & \cellcolor[HTML]{80ffb4} \textcolor[HTML]{000000}{3} & \cellcolor[HTML]{8000ff} \textcolor[HTML]{f1f1f1}{1} & \cellcolor[HTML]{000000} \textcolor[HTML]{f1f1f1}{0} & \cellcolor[HTML]{000000} \textcolor[HTML]{f1f1f1}{0} \\
\end{tabular}

    }
    \label{tab:cluster_class_matrices_dino_50}
\end{table*}

\subsubsection{Query Performance}
\begin{figure}
    \centering
    \includegraphics[width=0.95\columnwidth]{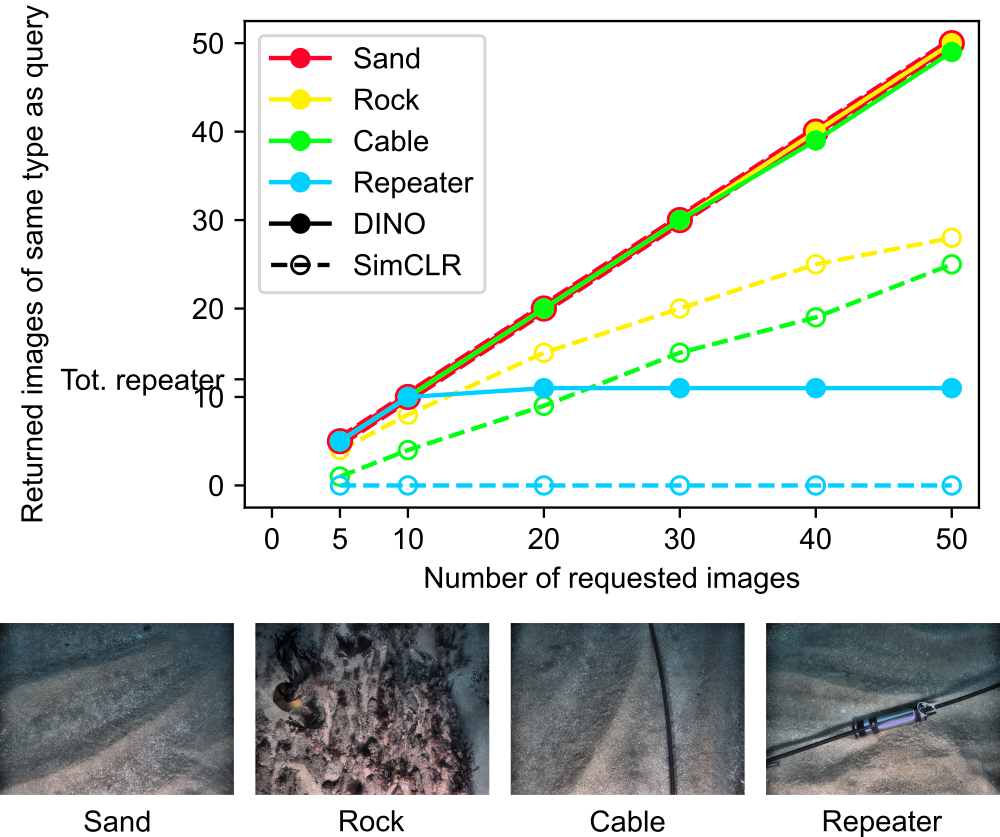}
    \caption{
        Number of returned images containing the same type of seafloor or objects as the query image vs. the number of requested images, when prompting remote query with any of the four images shown at the bottom.
        The marker and line style indicate whether the SimCLR or DINO model was used to encode the images.
    }
    \label{fig:query_analysis_result}
\end{figure}

The performance of remote queries was assessed by running the query algorithm using different query images and both models for encoding, and evaluating how many of the transmitted images matched the query image in terms of showing the same type of seafloor or objects on it.
The algorithm was tested with 4 different query images, selected from a previous dive at the same location.
Two of the images were representative of a ground truth class used in the summary evaluation (sand and rock), while the other two contained objects of interest (subsea communication cable and mock-up repeater device).
The algorithm was applied on the same dataset as for the summary evaluation, and was prompted to return the 5, 10, 20, 30, 40 and 50 most similar images to each query image, using both models.
The number of returned images that contained the type of seafloor or objects as the query image was counted, and the results are shown in Fig.~\ref{fig:query_analysis_result}.
As the images were not subsampled and the similarity metric is deterministic (i.e. no element of randomness involved in computing the cosine similarity of the image latents space representation), the results are repeatable for same settings, eliminating the need to measure variability across repeated runs.
For the query image showing sand, the algorithm returned images showing sand in 100\% of the cases, regardless of the model used or the tested number of requested images for transmission.
For the other query images the DINO model clearly outperformed the SimCLR model.
This stands out most clearly for when using the image of the mock-up repeater as prompt, where not a single image of the same type was returned when using the SimCLR encoder, while when using the DINO encoder, the algorithm returned (almost) all images of the repeater in the dataset. There are only 12 images of the mock-up repeater (full or partially) in the entire dataset, which is why the number of returned images of the same type as the query image plateaus when requested to return more than that.

Fig.~\ref{fig:remote_query} shows the results from the query for 50 images, using the image of the mock-up repeater device from Fig.~\ref{fig:query_analysis_result} as a prompt, and using the DINO model.

\subsubsection{Discussion}
The results show remote queries and summaries effectively summarise datasets of seafloor images, and can reliably find images similar to a query image.
While remote queries are inherently deterministic for a given dataset and query image, the results summary results contain a level of randomness, due to how the underlying algorithms work.
However, the results show to be robust in the face of this randomness.
Both algorithms also depend on the underlying encoder model used to extract image features.
In particular for remote queries, the DINO model showed a stronger performance than SimCLR, despite using the generic model.
On the other hand, SimCLR being a lighter model it can be trained using fewer resources than the DINO model and potentially be improved with more tailored training.

\section{Conclusion}
The method presented in this paper shows how information contained in seafloor image datasets can be compressed and transmitted over low-bandwidth communication links for over-horizon awareness of subsea imagery collected by AUVs.
The results from three successful campaigns demonstrate the method working at sea using different AUV platforms and seafloor mapping devices.
While it does not replace processing of the full data post-mission, it provides valuable information in near real-time to AUV operators about the imagery collected during a mapping mission and so can assist decision-making and remote retasking during over-horizon controlled AUV missions.
The ALR-BioCam deployment during the techoceans2024 campaign demonstrated an almost 400\,000-fold reduction in data volume compared to the raw data size, enabling transmission of data summaries of a 2-hour 47-minute-long mapping mission in just over 34 minutes over low-bandwidth satellite communication.
Despite the large reduction in data volume, semantic maps and representative images can show characteristic patterns of spatial distribution on the seafloor within tens of minutes of an AUV surfacing, without the need for recovery or proximity to the AUV.
Providing information about gathered data is one crucial step towards enabling AUVs to efficiently operate in long-range missions without a support vessel, and so reducing the cost and environmental impact of subsea surveys.

The method is flexible in that it can adapt to the capacity of the communication channel and the needs of the user or operational requirements, e.g. by adapting the size and number of transmitted compressed images, or by limiting the number or metadata sent.
This allows balancing the level of detail the transmitted data shows with the time available for transmitting data.
As the output simply is a number of files whose size can be defined beforehand, it makes the method easy to integrate with existing systems even if the mapping device does not have direct access to the acoustics or satellite communication modem. 
Data can be relayed through the AUV to the communication infrastructure as it does not need to understand the content of the packages and can treat them as black boxes.
While transmission via satellite or acoustic modems from AUVs is slow and, depending on sea conditions, can be intermittent, results can be shown as data packages arrive, even if they are received out of order.
Although dropped packages lead to gaps in the displayed data, it does not prevent the visualisation of the data that have been successfully transmitted, and so makes the method robust even with unreliable communication channels.

Wireless communication rates, such as satellite and underwater acoustics communication, have been increasing over the past years, and are expected to continue to do so in the future.
While this will allow to transmit larger amounts of data more quickly, the data rates achieved with compact, deep pressure-rated communication equipment used on AUVs are so far off from the data acquisition rates of imaging systems that it is unlikely that full datasets can be transmitted in real-time in the foreseeable future, and methods for compressing and summarising data will still be necessary.
Also, while the advantages of the demonstrated method are most obvious for scenarios with limited communication bandwidth, it is also beneficial in situations where higher bandwidth is available, as it provides summaries of entire datasets and is able to find images of interest automatically, and so is much faster than a human operator could sift through the data.

\section*{Acknowledgment}
The authors thank the staff from Oceanic Platform of the Canary Islands (PLOCAN) for their support during the techoceans2024 field trial, the crew of the survey vessel Echo Explorer from Sonardyne Intl. for their support during the EE25-01 field trial, and the Lerwick Port Authority, Phil Harris from Shetland Seabird Tours and the ALR Operations Team for their help during the OASIS ALR8 Shetland Trials 2025.


\bibliographystyle{IEEEtran}
\bibliography{bibliography}

\begin{IEEEbiography}[{\includegraphics[width=1in,height=1.25in,clip,keepaspectratio]{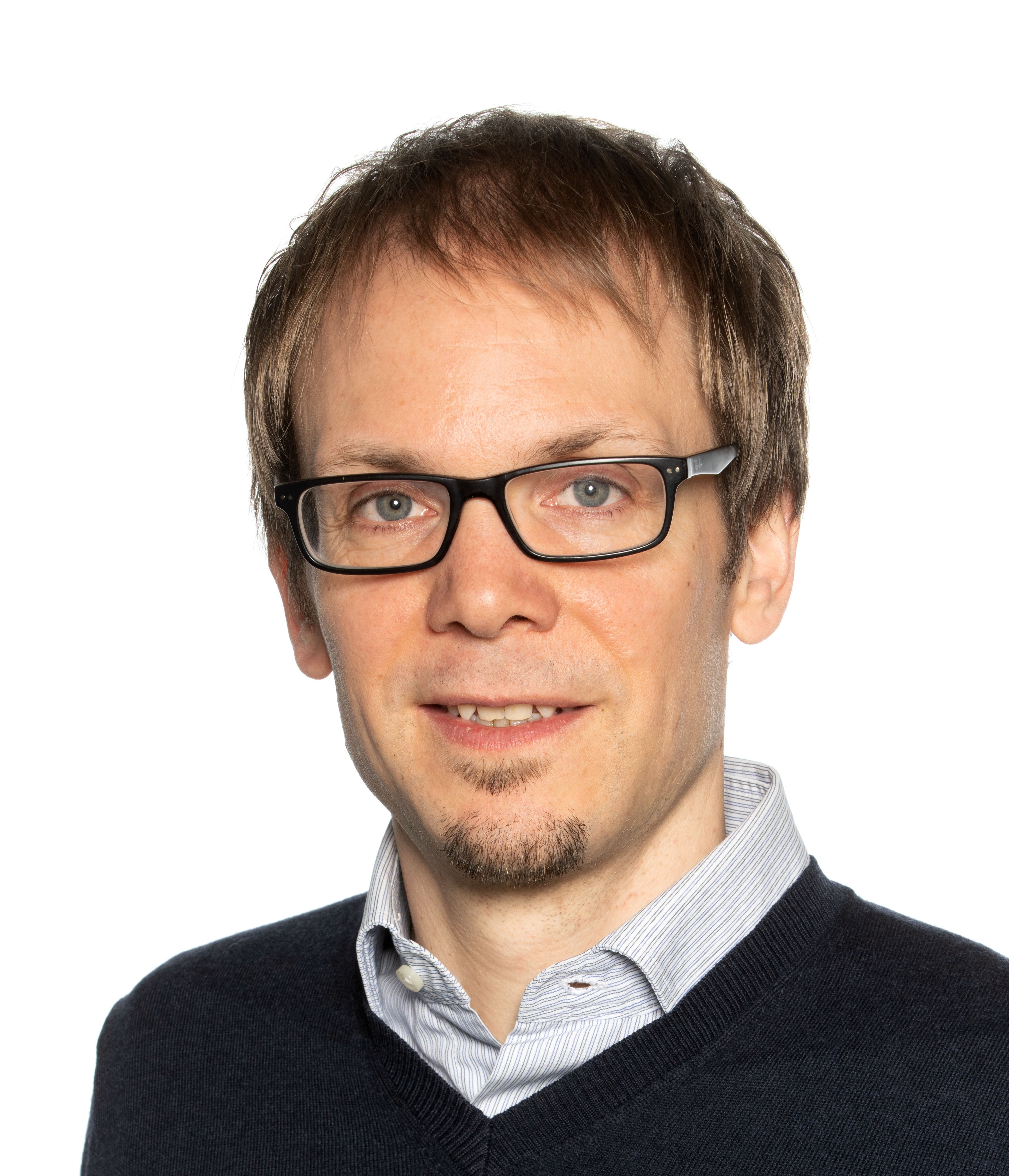}}]{Adrian Bodenmann} received a BSc in 2007 and a MSc in 2009 in Microengineering from the Ecole Polytechnique Fédérale de Lausanne (EPFL), Lausanne, Switzerland.
He worked at the Underwater Robotics and Applications (URA) laboratory at the Institute of Industrial Science at the University of Tokyo, Japan, as Project Researcher until 2017. He has since been working in the Maritime Engineering Group of the University of Southampton, Southampton, UK, as a Senior Research Assistant.
His research interests are centred around visual mapping of the seafloor, including development and deployment of deep-sea going mapping devices and development of algorithms for visual reconstruction and information extraction from seafloor imagery.
\end{IEEEbiography}

\begin{IEEEbiography}[{\includegraphics[width=1in,height=1.25in,clip,keepaspectratio]{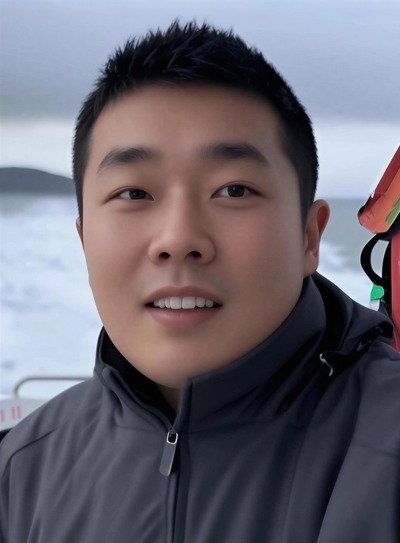}}]{Cailei Liang} received the BEng degree in navigation technology from Dalian Maritime University, China, in 2019. He subsequently obtained dual MSc degrees in traffic information engineering and control from Dalian Maritime University and Tokyo University of Marine Science and Technology, Japan, in 2022. He is currently pursuing a PhD degree in maritime engineering, University of Southampton, United Kingdom. His research interests include underwater robotic perception, computer vision for marine environments, and path planning for AUVs.
\end{IEEEbiography}

\begin{IEEEbiography}[{\includegraphics[width=1in,height=1.25in,clip,keepaspectratio]{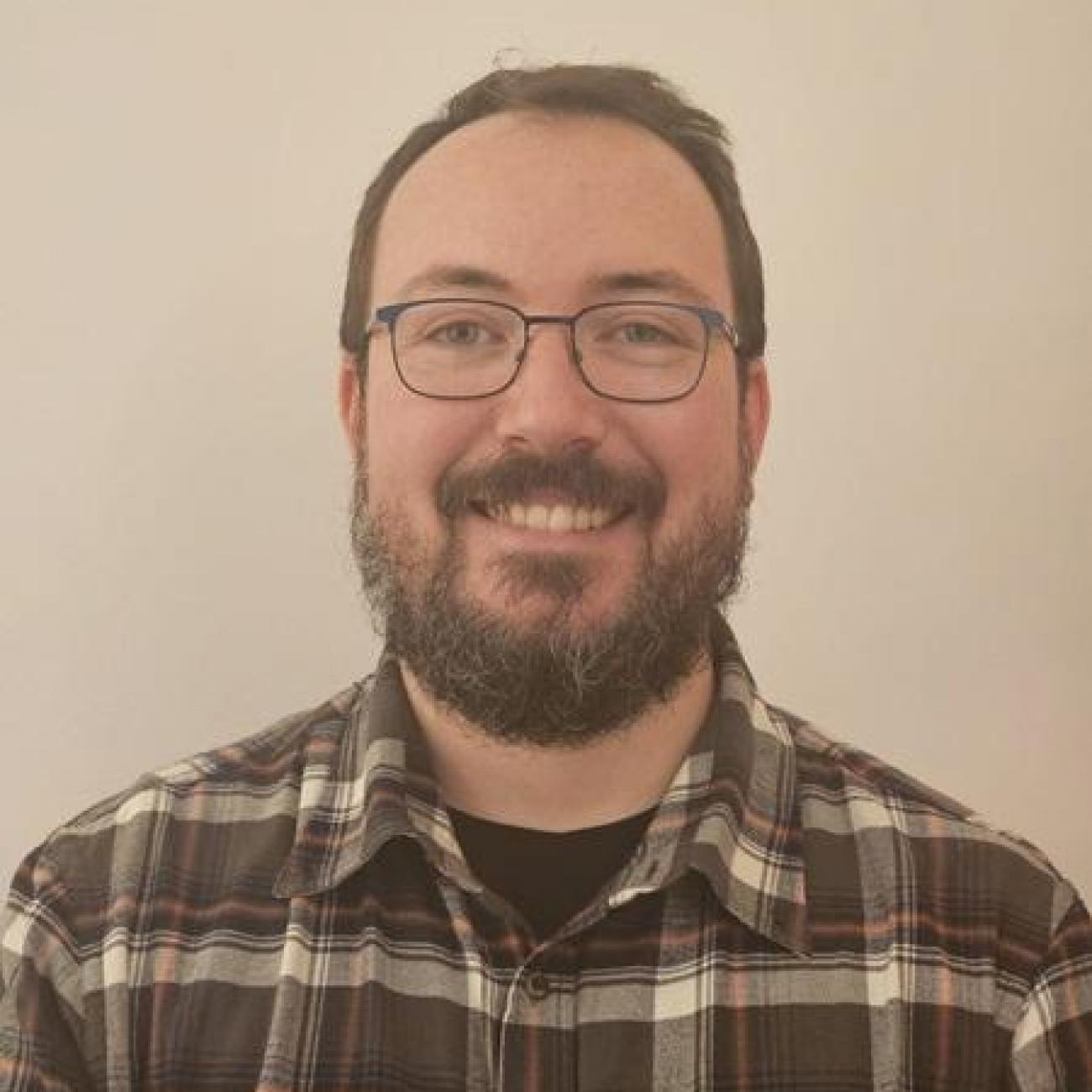}}]{Miquel Massot-Campos} was born in Palma de Mallorca, Spain. He received the MEng degree from Barcelona Tech (UPC) in 2011, the MSc degree from the University of the Balearic Islands in 2013, and the Ph.D. degree from the same institution in 2019.
From 2012 to 2018, he worked as a researcher in underwater robotics at SRV-UIB. From 2018 to 2024, he was a Senior Research Fellow at the University of Southampton, leading marine robotics and AI-driven seafloor survey research. Since 2024, he has been with Ocean Infinity, Southampton, UK, currently serving as Robotics Lead, overseeing a team of six engineers and mission-critical software deployment across 14 offshore vessels. His work has attracted over 1,400 citations, with an h-index of 19.
\end{IEEEbiography}

\begin{IEEEbiography}[{\includegraphics[width=1in,height=1.25in,clip,keepaspectratio]{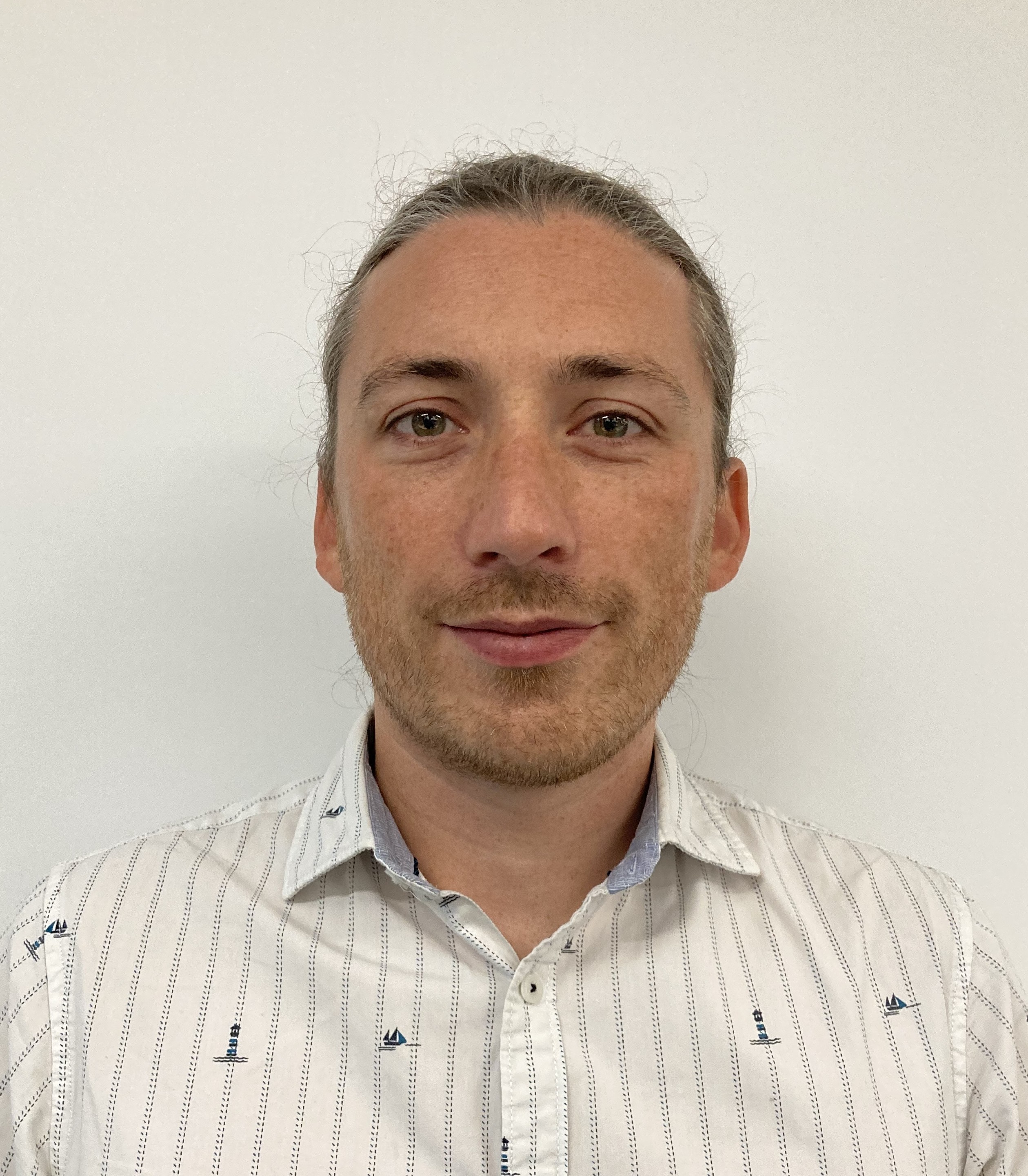}}]{Samuel Simmons} has worked in several fields including Maritime robotics, Flood Management and Acoustics. This has been supported with a Bachelors in Sound Design Technology from the University of the Hertfordshire, a Bachelors in Mechatronics from the University of Southern Denmark and a Foundation Degree in River and Coastal Engineering from the University of the West of England. Together these elements support his current role as a Maritime Robotics Engineer at the University of Southampton in the Maritime Engineering Group. His research interests are maritime robotics integration and development.
\end{IEEEbiography}

\begin{IEEEbiography}[{\includegraphics[width=1in,height=1.25in,clip,keepaspectratio]{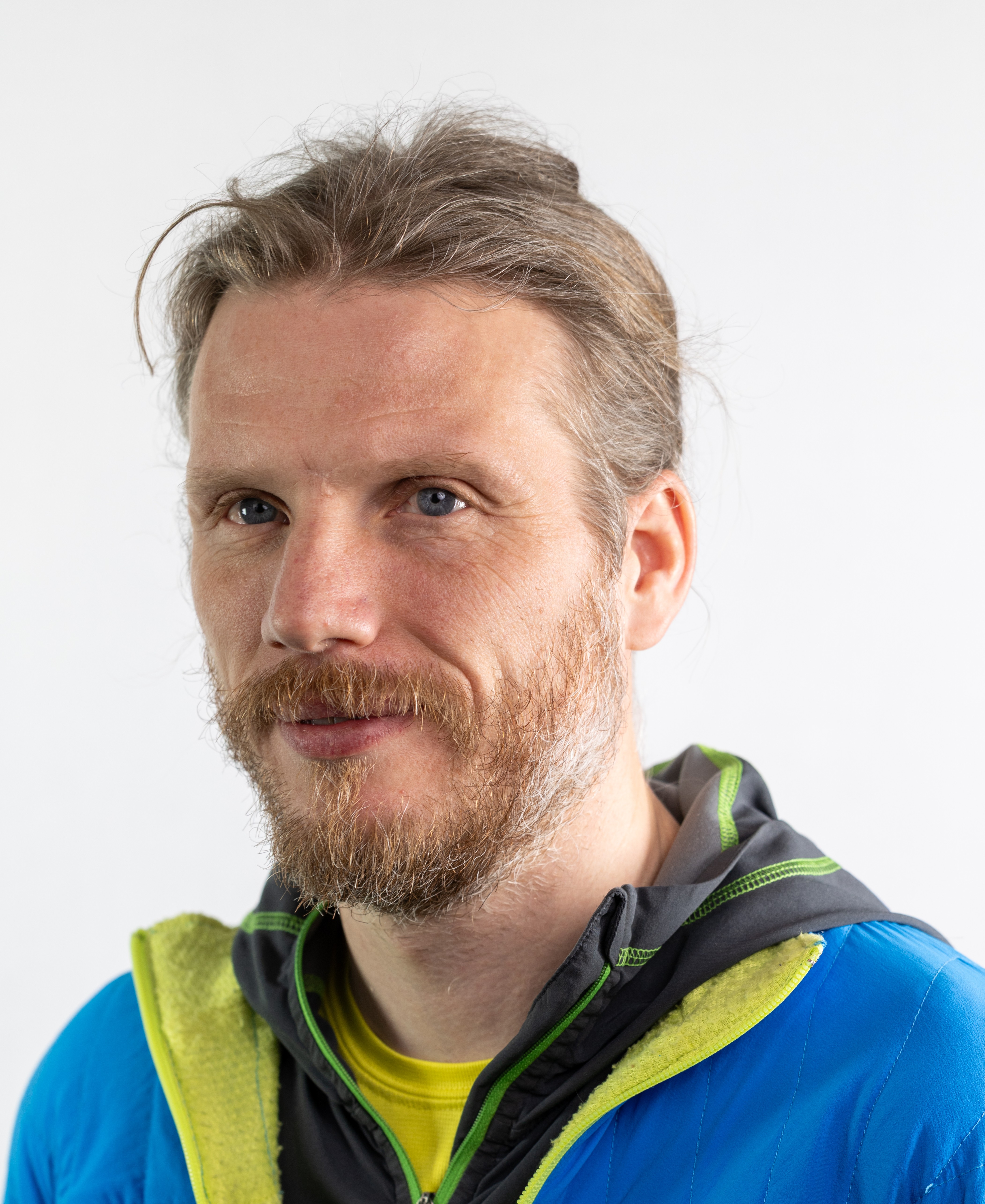}}]{Alexander B. Phillips} is Head of the Marine Autonomous and Robotic Systems (MARS) Group at the UK National Oceanography Centre (NOC), Southampton, UK. He received the MEng degree in Ship Science in 2004 and the Ph.D. degree in Maritime Robotics in 2010 from the University of Southampton, UK. At NOC, he leads a multidisciplinary team of approximately 70 engineers, researchers, and Ph.D. students engaged in the development and operation of advanced marine autonomous systems, including the Autosub family of AUVs. His research interests include marine robotics, AUVs, guidance, navigation and control, long-range autonomous operations, and intelligent systems for ocean observation. Dr. Phillips has coauthored more than 70 scientific publications, contributed to numerous national and international research and development projects, and led a wide range of field trials involving autonomous underwater systems.
\end{IEEEbiography}

\begin{IEEEbiography}[{\includegraphics[width=1in,height=1.25in,clip,keepaspectratio]{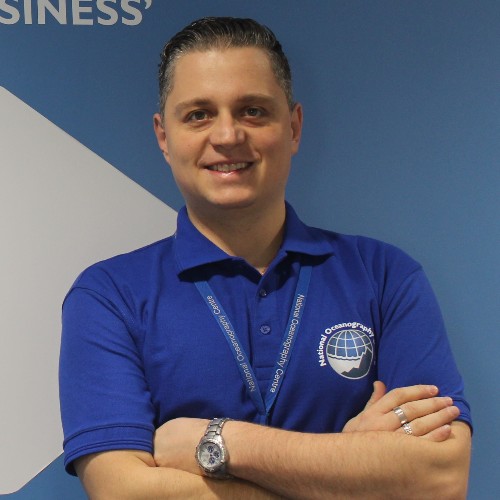}}]{Alberto Consensi} received the BEng degree in Electronics Engineering and the MSc degree in Robotics and Automation Engineering from the University of Pisa, Pisa, Italy, in 2011 and 2016, respectively. He also received the BA and MA degrees in Classical Music (Trumpet) from the Istituto Musicale P. Mascagni, Livorno, Italy, in 2005.
Since joining the National Oceanography Centre (NOC), Southampton, UK, in 2017, he has worked on the development of navigation algorithms, control, and sensor-integration for AUVs. He is currently a Senior Robotic Systems Engineer and Delivery Lead with the Marine Autonomous and Robotic Systems Group, where he leads projects focused on enabling autonomous capabilities for marine robotic systems, under-ice navigation, and long-range AUV operations. He is also Co-Lead for the NOC Arctic Network - Theme "Pioneering Technologies for Arctic Ocean Science".
Mr. Consensi is a Chartered Engineer with the Italian Engineering Council. His research interests include AUVs, marine robotic systems, navigation and localization, multi-sensor fusion, and under-ice autonomy.
\end{IEEEbiography}

\begin{IEEEbiography}[{\includegraphics[width=1in,height=1.25in,clip,keepaspectratio]{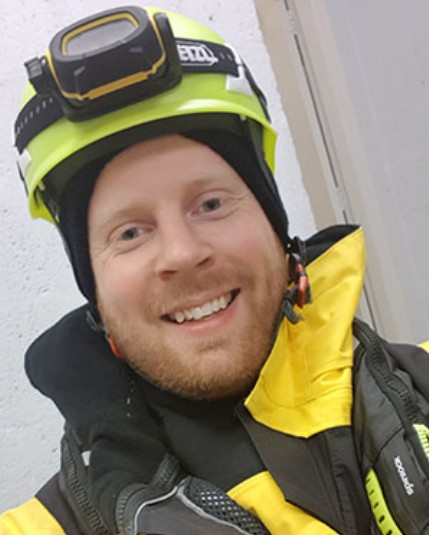}}]{Matthew Kingsland} received the BEng degree in systems and electronics engineering from Loughborough University, Loughborough, UK
He has been in the marine industry since 2012 and joined the National Oceanography Centre, Southampton, UK, in 2018, as a Senior Robotics Systems Engineer. 
His research interests include AUV systems architecture, under ice AUVs, and subsea power systems.
\end{IEEEbiography}

\begin{IEEEbiography}[{\includegraphics[width=1in,height=1.25in,clip,keepaspectratio]{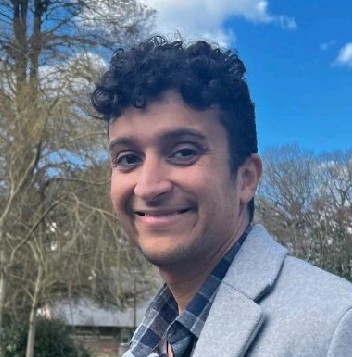}}]{Rashiid Sherif} received a BEng in Electrical and Electronics Engineering in 2016 and a MEng in Embedded microelectronics and Wireless Systems in 2017 from Coventry University, Coventry, UK. He subsequently joined KROHNE Group, where he worked as a Manufacturing Systems Development Engineer where he worked until 2019 and James Fisher Prolec, where he worked as a Hardware Engineer until 2020. He then joined the National Oceanography Centre Southampton, where eh worked as a Robotic System Software Engineer, working on the embedded software of the Autosub AUV fleet, before joining Ocean Infinity as a Robotics Software Engineer in 2025.
\end{IEEEbiography}

\begin{IEEEbiography}[{\includegraphics[width=1in,height=1.25in,clip,keepaspectratio]{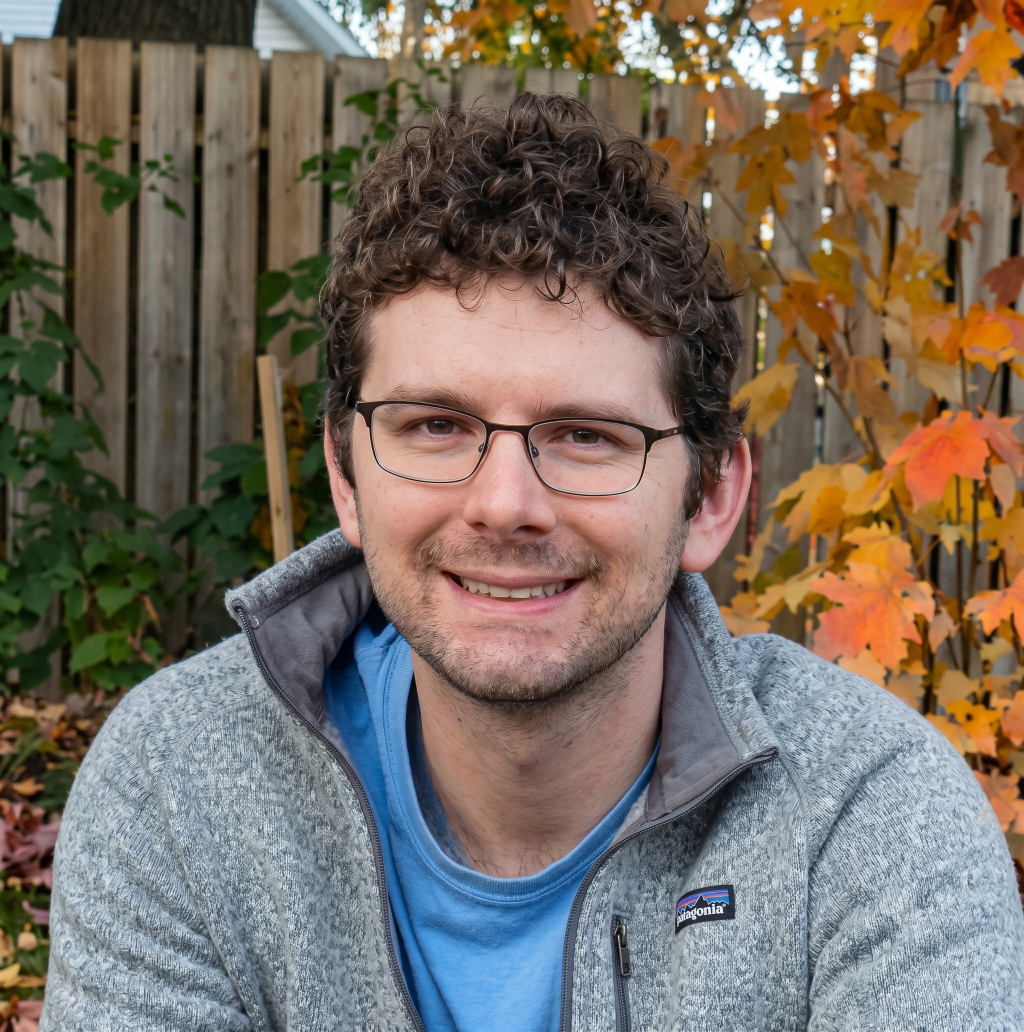}}]{Stanley Brown} received his BSc in Environmental Engineering from the University of British Columbia, Canada, in 2012, and his MSc from the University of Waterloo, Canada, in 2018. He subsequently worked as a Software and Computer Vision Engineer at Voyis, focusing on underwater stereo camera systems, photogrammetry, calibration, and applying machine learning to challenging underwater datasets. His core research interests include computer vision, robotics, state estimation, and calibration.
\end{IEEEbiography}

\begin{IEEEbiography}[{\includegraphics[width=1in,height=1.25in,clip,keepaspectratio]{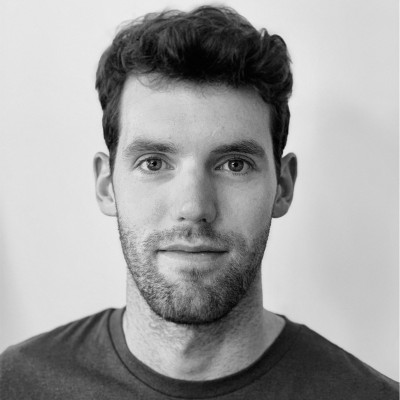}}]{Adam Riese} is originally from Winnipeg, Canada. He earned a Bachelor of Applied Science in Nanotechnology Engineering from University of Waterloo (2013) and a Master of Applied Science in Mechanical \& Materials Engineering from Western University in London, Canada (2015). Following a couple of years in the satellite industry, Adam joined Voyis (then 2G Robotics) in 2017 and held various roles in manufacturing and engineering before becoming VP, Engineering in 2020. In his current role he is responsible for the development of Voyis underwater optical sensors and embedded systems. He is a registered Professional Engineer in Ontario.
\end{IEEEbiography}

\begin{IEEEbiography}[{\includegraphics[width=1in,height=1.25in,clip,keepaspectratio]{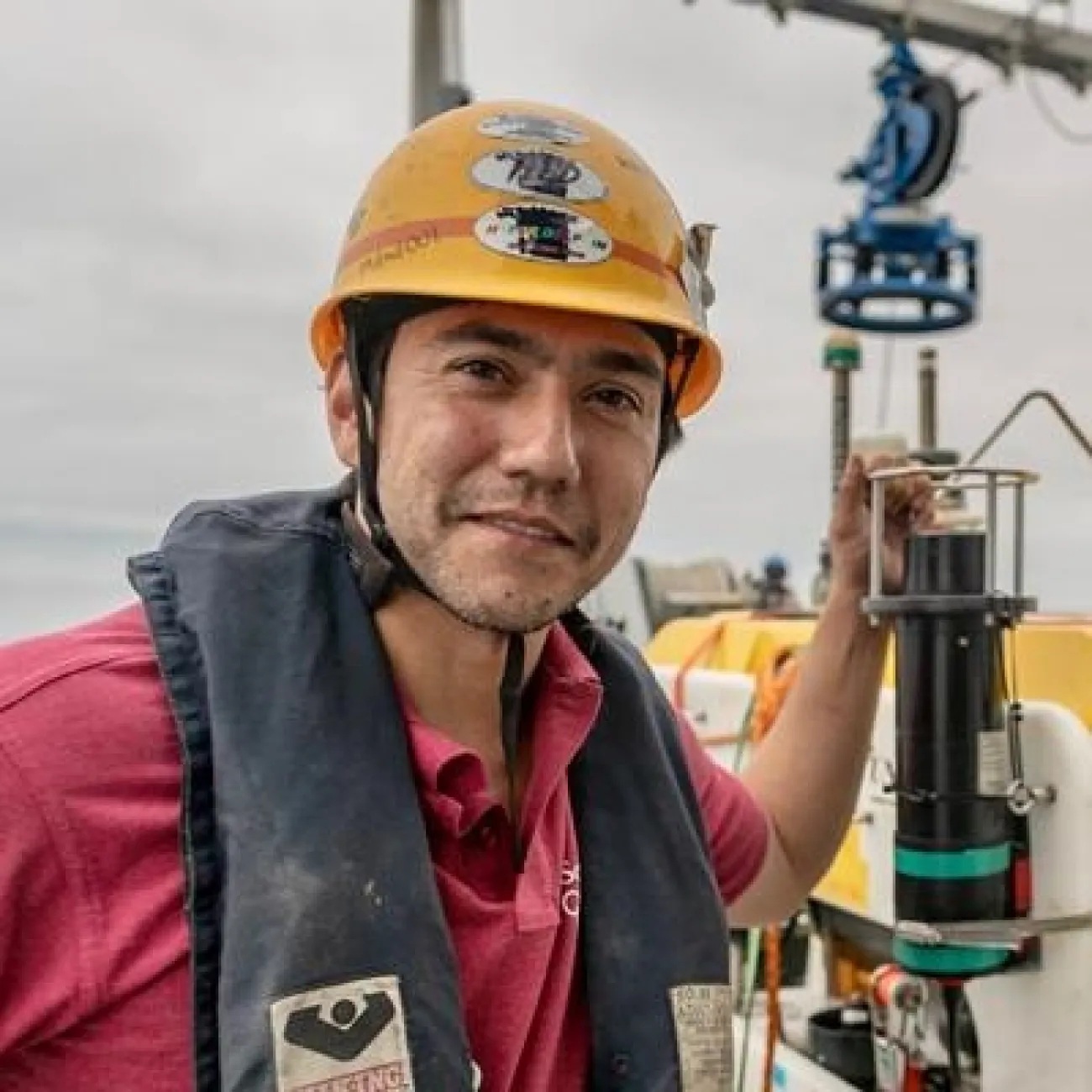}}]{Blair Thornton} is Professor of Marine Autonomy at the University of Southampton, UK, with a cross-appointment at the Institute of Industrial Science, University of Tokyo, Japan. He has over 20 years of international experience developing autonomous robotic submersibles, sensors, and algorithms for intelligent navigation, data processing, and scalable operation. He has spent more than 500 days at sea on over 50 expeditions, deploying robotic systems he has developed in support of marine science and statutory monitoring, including applications in conservation, deep-sea exploration, infrastructure inspection, and disaster response. His research focuses on addressing bottlenecks in the flow of information from data collection to human insight and decision support.
\end{IEEEbiography}

\end{document}